\documentclass[11pt]{article}

\usepackage{amsmath}
\usepackage{amsfonts}
\usepackage{mathrsfs}

\usepackage[nohead,letterpaper]{geometry}
\usepackage{vmargin}
\setpapersize{USletter}
\setmargrb{2cm}{1cm}{2cm}{1.75cm}

\usepackage{xcolor}
\definecolor{plum}{rgb}{0.36078, 0.20784, 0.4}
\definecolor{chameleon}{rgb}{0.30588, 0.60392, 0.023529}
\definecolor{cornflower}{rgb}{0.12549, 0.29020, 0.52941}
\definecolor{scarlet}{rgb}{0.8, 0, 0}
\definecolor{brick}{rgb}{0.64314, 0, 0}

\newcommand{\ts}[1]{\textrm{\tiny #1}}
\newcommand{\ms}[1]{\textrm{\tiny $#1$}}

\newcommand{\email}[1]{\href{mailto:#1}{\tt \textcolor{cornflower}{#1}}}

\newcommand\abig{\mbox{\Large\textit{a}}}

\newcommand{\dub}{\,\,\,\,}

\newcommand\MM{\mathcal{M}}
\newcommand\RR{\mathcal{R}}

\newcommand\OO{\mathcal{O}}
\newcommand\DD{\mathcal{D}}
\newcommand\CC{\mathcal{C}}
\newcommand\EE{\mathcal{E}}

\newcommand\dM{\partial \MM}
\newcommand\FF{\mathcal{F}}

\newcommand\LL{\mathcal{L}}

\newcommand\eps{\epsilon}
\newcommand\nts{\negthickspace}
\newcommand\bns{\nts \nts \nts}
\newcommand\hK{\hat{K}}
\newcommand\defeq{\mathrel{\mathop:}=}
\newcommand{\BR}{\mathbb{R}}

\usepackage[%
  pdftitle={Boundary Terms Unbound!},%
  pdfauthor={Robert Mann, Robert McNees},%
  pdfsubject={Holographic Renormalization},%
  pdfkeywords={Holographic Renormalization, Boundary Terms, Surface Terms, Variational 
    Principle, NS 5-branes, Linear Dilaton, Asymptotically Linear Dilaton},%
  colorlinks=true,linkcolor=cornflower,citecolor=brick,urlcolor=chameleon%
]{hyperref}

\numberwithin{equation}{section}

\usepackage[bf,medium,]{titlesec}
\titleformat{\section}
{\large\bfseries}{\thesection.}{.5em}{\large\bfseries}

\titleformat{\subsection}
{\bfseries}{\thesubsection}{.5em}{\bfseries}

\begin{document}


\thispagestyle{empty}

\begin{flushright}
arXiv:0905.3848 \\ \vskip.5em
\today
\end{flushright}
~\vspace{1.75cm}\\

\begin{center}
{\bf \Large Boundary Terms Unbound!}\vskip.5em
         {\Large Holographic Renormalization of \\ Asymptotically Linear Dilaton Gravity.
}
\end{center}

\vspace{.5cm}

\begin{center}
Robert B. Mann $^{a, b}$ and Robert McNees $^{b}$

\vspace{1cm}{\small {\textit{$^{a}$Department of Physics,\\ University of
Waterloo,\\ Waterloo, Ontario N2L 3G1,\\ Canada}}}\\
\vspace{2mm} {\small {\textit{$^{b}$Perimeter Institute for Theoretical Physics,\\ 31 Caroline Street North,\\ Waterloo,
Ontario N2L 2Y5,\\ Canada}}}\\
\vspace*{0.5cm}
\end{center}
\vspace{0.5cm}

\begin{center}
{\bf Abstract}
\end{center}

A variational principle is constructed for gravity coupled to an asymptotically linear dilaton and a $p$-form field strength. This requires the introduction of appropriate surface terms -- also known as `boundary counterterms' -- in the action. The variation of the action with respect to the boundary metric yields a boundary stress tensor, which is used to construct conserved charges that generate the asymptotic symmetries of the theory. In most cases a minimal set of assumptions leads to a unique set of counterterms. However, for certain examples we find families of actions that depend on one or more continuous parameters. We show that the conserved charges and the value of the on-shell action are always independent of these parameters.

\vfill
\noindent\makebox[5cm][l]{\hrulefill}\\
{\footnotesize \email{rbmann@sciborg.uwaterloo.ca}\\
			\email{rmcnees@perimeterinstitute.ca}}\\
\newpage

\setcounter{page}{1}

\tableofcontents


\section{Introduction}
\label{sec:Intro}

Holographic renormalization was first understood as an important ingredient in gauge/gravity dualities \cite{Henningson:1998gx,Balasubramanian:1999re,Emparan:1999pm,Bianchi:2001kw}. In that context, the goal is to address a class of large volume (IR) divergences in the gravitational action. This is accomplished by adding a number of surface terms that render the action and other quantities of interest finite. Soon afterward, the technique was extended to theories with qualitatively different spacetime asymptotics, like de Sitter space \cite{Balasubramanian:2001nb,Balasubramanian:2002zh,GM2002}, and inflating FRW spacetimes \cite{Larsen:2003pf, Larsen:2004kf}. This marked a new application of the technique, as such theories are not known to possess a holographic dual. It also makes ``holographic renormalization" something of a misnomer, but the term has stuck and so we will use it.

Before the development of holographic renormalization, the problem of IR divergences in calculations involving gravitational actions was usually addressed via a technique known as `background subtraction' \cite{Gibbons:1976ue}. Given a spacetime $(\MM,g)$, one attempts to isometrically embed a regulating boundary $(\dM,h)$ in some suitable reference spacetime $(\MM_{\textrm{\tiny ref}}, g_{\textrm{\tiny ref}})$ \cite{Brown:1992br, Brown:2000dz}. The action is then obtained from the difference of the regulated actions for $(\MM,g)$ and $(\MM_{\textrm{\tiny ref}}, g_{\textrm{\tiny ref}})$, in the limit that the regulating boundary is removed to infinity. In many calculations this technique produces a physically sensible result. Indeed, several important results in Euclidean quantum gravity are based on these sorts of calculations. However, background subtraction also suffers from a number of problems \cite{Emparan:1999pm, Mann:2005yr}. For example, there may be more than one candidate for the reference spacetime, and these may lead to inequivalent results. Furthermore, in spacetime dimension $d+1 \geq 4$, Weyl's embedding theorem states that an isometric embedding of the regulating boundary in the reference spacetime may not exist. In other words, there is no guarantee that background subtraction can be implemented for a particular spacetime. These problems are not restricted to exotic or pathological spacetimes; even well-known solutions like the Kerr spacetime suffer from ambiguities \cite{Martinez:1994ja}. Holographic renormalization, on the other hand, is a completely intrinsic technique that does not rely on reference spacetimes or particular embeddings.

There is another motivation for holographic renormalization that is entirely separate from gauge/gravity dualities. In the 1970s, Regge and Teitelboim \cite{Regge:1974zd} showed that gravity with asymptotically flat boundary conditions requires additional surface terms in order for the Hamiltonian to have the appropriate variational properties. The same is true for the Lagrangian approach: in non-compact spacetimes, physically interesting boundary conditions for the metric \emph{require} additional surface terms in the action. In fact, the AdS boundary counterterms serve precisely this purpose \cite{Papadimitriou:2005ii}. The IR divergences that plague the standard form of the action are symptomatic of a deeper issue: the lack of a well-defined variational principle. This is a fundamental problem that background subtraction cannot address. Even if the regulating boundary for a spacetime can be embedded in an appropriate reference background, such embeddings cannot be constructed for an open family around that spacetime \cite{Mann:2005yr}. This means that background subtraction cannot be used -- even in principle -- to construct a well-defined variational principle. Holographic renormalization offers a complete and unambiguous solution to this problem.

Whether we are discussing gauge/gravity dualities or Euclidean quantum gravity, it is essential to have a well-defined variational principle. This is the key physical insight that guides the introduction of surface terms in the action, as well as the derivation and interpretation of conserved charges. The variational approach to holographic renormalization is especially relevant for gravity with asymptotically flat boundary conditions. Early results for AAdS spacetimes make use of the constant-curvature asymptotics, and hence do not generalize to asymptotically flat spacetimes (though some progress was made in \cite{Kraus:1999di}). A different approach, which makes use of the Gauss-Codazzi relations, was recently proposed for gravity with asymptotically flat boundary conditions \cite{Mann:2005yr}. The new surface term is proportional to the trace of a tensor $\hK_{ab}$, which is constructed from the intrinsic curvature of the induced metric on the (regulating) boundary. The resulting action
 \begin{equation}
 \label{covaction} I = \frac{1}{2\,\kappa^{\,2}}\int_{\MM} \nts d^{\,d+1}x \,\sqrt{-g}\,R +
 \frac{1}{\kappa^{\,2}}\int_{\dM} \bns d^{\,d}x \, \sqrt{-h} \,\big(K - \hat K\big),
 \end{equation}
is finite on-shell, and its first variation vanishes for all asymptotically flat variations of the metric. Note that ``asymptotically flat variation" has a precise meaning here: it refers to a specific set of fall-off conditions on components of the metric, as well as certain kinematic constraints on the metric at spatial infinity \cite{Ashtekar:1990gc, Ashtekar:1991vb, Beig:1980be, Beig:1982xx}. The boundary stress tensor obtained from the variation of this action is proportional to the electric part of the Weyl tensor, and it can be shown \cite{Mann:2008ay} that currents built from the stress tensor can be integrated over a cut of spatial infinity to yield conserved charges that are equivalent to both the ADM construction \cite{ADM1} and the covariant charges defined by Ashtekar and Hansen \cite{AH}.

For gravity with asymptotically flat boundary conditions the ``$\hK$" counterterm prescription \cite{Mann:2005yr, Mann:2006bd, Astefanesei:2006zd, Mann:2008ay} has been shown to be quite robust. It has also been applied to a broad class of spacetimes obeying these boundary conditions \cite{Ross2008}. This was also the case with the original AdS/CFT construction for boundary counterterms, which applies to all spacetimes with AAdS boundary conditions \cite{Emparan:1999pm, Papadimitriou:2005ii}. So it is a bit odd that the two constructions seem to be incompatible. That is, the $\hK$ approach does not work for AAdS boundary conditions, and the AdS/CFT constructions do not work for AF boundary conditions. To investigate this situation, we turn to a recent application of the $\hK$ construction by Marolf and Virmani \cite{Marolf:2007ys}, who consider NS 5-brane backgrounds in ten dimensional type II supergravity. They develop a hybrid approach which depends on a $\hK$ surface term, as well as additional surface terms reminiscent of those used in AdS/CFT. Motivated by their example, we consider a broader class of theories in which gravity is coupled to an asymptotically linear dilaton field and magnetic $p$-form flux. Because of the relation to the NS 5-brane background, we refer to this class of theories as asymptotically linear dilaton (ALD) gravity. In some respects, these theories are similar to the more familiar AF and AAdS cases. But obtaining the proper surface terms reveals new aspects of holographic renormalization. For instance, we find that certain ALD theories allow for a family of actions that admit a well-defined variational principle, have the same value on-shell, and yield the same conserved charges. These results add to existing work on non-conformal brane backgrounds \cite{Cai:1999xg, Kanitscheider:2008kd, Wiseman:2008qa}.

The outline of our paper is as follows. We begin in section \ref{sec:Prelim} by defining ALD gravity and showing that the usual form of the action does not give a well-defined variational principle. In section \ref{sec:KHat} we discuss how the $\hK$ prescription can be applied to ALD gravity, and then examine a broader range of possible boundary counterterms for this class of theories. We continue along these lines in section \ref{sec:OC}, where we offer an alternative prescription for the holographic renormalization of ALD gravity.  In section \ref{sec:LRF} we compute the boundary stress tensor and the associated conserved charges.  In section \ref{sec:Examples} we apply our results to two examples in ten dimensional, type II supergravity: spacetimes asymptotic to a stack of NS 5-branes, and spacetimes asymptotic to a stack of D5-branes.  We summarize our results in section \ref{sec:Discussion}, and present technical details of our calculations in a collection of Appendices. 

Throughout most of the paper we set $2\,\kappa^{\,2} = 16 \pi G = 1$, but these factors are restored for the examples in section \ref{sec:Examples}. Furthermore, the dimension of spacetime is taken to be $d+1 > 4$. This assumption is meant to expedite our calculations, which would be complicated in lower dimensions by the mixing of terms at different orders in the asymptotic expansions. Our expectation, based on a similar situation in \cite{Mann:2008ay}, is that the results of sections \ref{sec:KHat}, \ref{sec:OC}, and \ref{sec:LRF} readily generalize to the case $d+1=4$.

\section{Asymptotically Linear Dilaton Gravity}
\label{sec:Prelim}

This section is divided into four parts. In the first part we give the action for gravity coupled to a dilaton and field strength on a manifold with boundary. In the second part we consider the same theory on a spacetime with non-compact spatial slices, and define a set of asymptotically linear dilaton (ALD) boundary conditions that characterize the behavior of the fields near spatial infinity. The third part gives a prescription for working with functionals that are ill-defined because of the asymptotic behavior of the fields. Finally, in the last part we show that solutions of the equations of motion with ALD boundary conditions are not stationary points of the original action.

\subsection{Equations of Motion}
\label{sec:EOM1}

We begin by considering gravity coupled to a scalar field and a $p-1$ form ${\bf A}$, on a $d+1$ dimensional spacetime $(\MM,g)$ with boundary $(\dM,h)$. This theory is described by the Einstein frame action
\begin{equation}\label{bulkaction}
  I = \int_{\MM} \nts \nts d^{\,d+1}x \sqrt{g} \, \left( R - \frac{4}{d-1} \, \nabla^{\mu} \phi
    \nabla_{\mu} \phi - \frac{1}{2 \cdot p!} \, e^{2 \alpha \phi} \, F^{a_1 \ldots a_p}
    F_{a_1 \ldots a_p}\right) +  2 \int_{\dM} \bns \nts d^{\,d} x \,\sqrt{h} \, K
\end{equation}
where ${\bf F} = {\rm d} {\bf A}$, and $K$ is the trace of the extrinsic curvature of the boundary. Under a small variation of the fields $(\delta g_{ab}, \delta \phi, \delta {\bf A})$, with the restriction to $\dM$ denoted by $(\delta h_{ab}, \delta \varphi, \delta {\bf \abig})$, the change in the action is given by
\begin{align}\label{ActionVariation1}
 \delta I = & \int_{\MM} \nts \nts d^{\,d+1}x \sqrt{g} \, \left( \EE^{ab} \, \delta g_{ab}
 + \EE_{\phi} \, \delta \phi + \EE^{a_1 \ldots a_{p-1}} \, \delta A_{a_1 \ldots a_{p-1}}  \right) \\ \nonumber
 &  +  \int_{\dM} \bns \nts d^{\,d} x \, \sqrt{h} \, \left( \pi^{ab} \, \delta h_{ab}
 + \pi_{\phi} \, \delta \varphi + \pi^{a_1 \ldots a_{p-1}} \, \delta \abig_{a_1 \ldots a_{p-1}} \right) ~.
\end{align}
The coefficients of the field variations in the bulk integral, when set to zero, yield the equations of motion of the theory
\begin{gather}\label{EinsteinEOM}
 R_{ab} - \frac{1}{2}\,g_{ab}\, R = \frac{4}{d-1}\,\left( \nabla_{a} \phi \nabla_{b}
    \phi -\frac{1}{2}\,g_{ab}\,(\nabla \phi)^2 \right)   + \frac{1}{2 \, (p-1)!}\,e^{2\alpha \phi} \, \left(F_{a}{}^{\ldots}\,F_{b \ldots } - \frac{1}{2\,p}\,g_{ab}\,F^{\,2}\right)\\\label{DilatonEOM}
 \nabla^{2} \phi - \frac{\alpha \, (d-1)}{8\,p!}\, e^{2 \alpha \phi} \, F^{\,2} = 0 \\ \label{MaxwellEOM}
 \nabla_{a_1} \left(e^{2\alpha \phi} \, F^{a_1 \ldots a_p} \right) = 0~.
\end{gather}
The boundary terms that remain in \eqref{ActionVariation1} have the form $p\,\delta q$, where $\delta q$ is a field variation on $\dM$ and $p$ is the momentum conjugate to $q$. As long as these momenta are finite at $\dM$, the boundary terms in $\delta I$ will vanish if the fields satisfy Dirichlet boundary conditions.

This derivation of the equations of motion assumes that spacetime has a boundary where the field variations vanish. But we are often interested in spacetimes with non-compact spatial slices, where instead of a boundary one is interested in the behavior of the fields at spatial infinity. In that case these equations do not give stationary points of the action \eqref{bulkaction}. This is not a surprising result: gravity with asymptotically flat boundary conditions suffers from the same problem \cite{Regge:1974zd, Mann:2005yr, Mann:2006bd}. The main goal of this paper is to construct an action with a well-defined variational principle that describes asymptotically linear dilaton gravity on spacetimes with non-compact spatial slices. Our first assumption regarding this action is that its stationary points should satisfy the equations \eqref{EinsteinEOM}-\eqref{MaxwellEOM}.

\subsection{ALD Boundary Conditions}
\label{sec:BndyCond}

For spacetimes with non-compact spatial slices, the variational principle must distinguish between different field configurations that exhibit the same behavior near spatial infinity. This requires that we set fall-off conditions for the fields along some spacelike coordinate $\rho$, with spatial infinity at $\rho \to \infty$, and determine appropriate kinematic constraints on the asymptotic values of the fields. In a slight abuse of terminology we will refer to these specifications as ``boundary conditions".

\vskip1em
\noindent{\bf Fall-off Conditions}
\vskip.5em

\noindent The fall-off conditions are expressed in terms of asymptotic expansions of the fields. For $\rho \gg 1$ the line element takes the form
\begin{gather}\label{metricBC}
  ds^{2} = d\rho^{2} + \rho^{2} \, \left(h_{ab}^\ts{(0)} + \rho^{2-d} \,
    h_{ab}^\ts{(1)} + \rho^{1-d} \, h_{ab}^\ts{(2)} + \ldots \right)\,dx^{a} dx^{b} ~.
\end{gather}
Let $\Sigma$ denote a $d$-dimensional surface of constant $\rho$\,; then the leading term $h_{ab}^\ts{(0)}$ is a metric on $\Sigma$, and the $h_{ab}^\ts{(n)}$ are smooth, symmetric tensor fields on $\Sigma$. The properties of constant $\rho$ surfaces are reviewed in appendix \ref{sec:Hyper}. We assume that $\Sigma$ consists of a timelike component $\Sigma_\ts{A}$ with dimension $d-p$ and a spacelike component $\Sigma_\ts{B}$ with dimension $p$
\begin{gather}
  \Sigma \sim \Sigma_\ts{A} \times \Sigma_\ts{B} ~.
\end{gather}
The component $\Sigma_\ts{A}$ has coordinates $x^{i}$ and metric $h_{ij}^\ts{(0)}$, and the component $\Sigma_\ts{B}$ has coordinates $x^{\alpha}$ and metric $h_{\alpha\beta}^\ts{(0)}$.

The asymptotic expansion of the dilaton is similar to \eqref{metricBC}, except the leading term is logarithmic
\begin{align}\label{dilatonBC}
    \phi = & \,\,\bar{\phi} \,\log(\rho) + \phi^\ts{(0)} + \rho^{\,2-d}\,\phi^\ts{(1)} + \rho^{\,1-d}\,\phi^\ts{(2)} + \ldots ~.
\end{align}
The quantities $\bar{\phi}$ and $\phi^\ts{(0)}$ are constants, and the remaining $\phi^\ts{(n)}$ are smooth functions on $\Sigma$. Instead of specifying fall-off conditions for the $p-1$ form ${\bf A}$, it is more convenient to work directly with the field strength ${\bf F}$. The expansion for the field strength is
\begin{align} \label{fluxBC}
    F_{a_1 \ldots a_p} = & F_{a_1 \ldots a_p}^\ts{(0)} + \rho^{\,2-d} \, F_{a_1 \ldots a_p}^\ts{(1)} + \rho^{\,1-d} \, F_{a_1 \ldots a_p}^\ts{(2)} \\ \nonumber
    & \, \, + \rho^{\,1-d} \, n_{{\textrm{\tiny [}}\,a_1} \, t^\ts{(1)}_{a_2 \ldots a_p \, \textrm{\tiny ]}}
    + \rho^{-d} \, n_{{\textrm{\tiny [}}\,a_1} \, t^\ts{(2)}_{a_2 \ldots a_p \, \textrm{\tiny ]}} + \ldots ~.
\end{align}
The leading term in this expansion is assumed to have support only on the $p$-dimensional subspace $\Sigma_{B}$, i.e.
\begin{gather}
F_{\alpha_1 \ldots \alpha_p}^\ts{(0)} = Q \, \eps_{\alpha_1 \ldots \alpha_p}^\ms{(0)}~,
\end{gather}
with all other components at this order equal to zero. The remaining terms are smooth, anti-symmetric tensors on $\Sigma$, with $n^{a}$ the spacelike unit vector normal to $\Sigma$.

\vskip1em
\noindent{\bf Boundary Data}
\vskip.5em

\noindent The leading terms in the asymptotic expansions \eqref{metricBC}, \eqref{dilatonBC}, and \eqref{fluxBC} constitute the boundary data for the fields. In principle one may specify any boundary data and consider the set of all field configurations that satisfy the resulting boundary conditions. However, this set will only contain solutions of the equations of motion if the boundary data satisfies the equations obtained from the leading terms in the asymptotic expansions of \eqref{EinsteinEOM}-\eqref{MaxwellEOM}. These kinematic constraints must therefore be imposed as part of our boundary conditions.

The leading term in the dilaton equation \eqref{DilatonEOM} fixes $\bar{\phi}$ and $\phi^{_{(0)}}$ in terms of other constants in the theory. If we define the parameter $\beta$ as
\begin{equation}
  \beta^{\,2\,(p-1)} \defeq \frac{8\,(p-1)}{\alpha^{2}\,Q^{2}}~,
\end{equation}
then the first two terms in \eqref{dilatonBC} are required to take the form
\begin{gather}\label{DilatonConstraint}
 \bar{\phi} \,\log(\rho) + \phi^\ts{(0)} = \frac{p-1}{\alpha}\,\log(\beta\,\rho) ~.
\end{gather}
The leading term in the Einstein equation \eqref{EinsteinEOM} fixes the Ricci tensor of the $d$-dimensional metric $h_{ab}^\ts{(0)}$. The components of $\RR_{ab}^\ts{(0)}$ on $\Sigma_\ts{A}$ and $\Sigma_\ts{B}$ are
\begin{gather} \label{Ricci2}
  \RR_{ij}^\ts{(0)} = (d-1)\,\left(1-\frac{4\,(p-1)^2}{\alpha^{2}\,(d-1)^2} \right)\,
    h_{ij}^\ts{(0)} \\ \label{Ricci1}
  \RR_{\alpha\beta}^\ts{(0)} = (d-1)\,\left( 1+ \frac{4\,(p-1)\,(d-p)}{\alpha^{2}\,(d-1)^{2}} \right)\,h_{\alpha\beta}^\ts{(0)} ~.
\end{gather}
For convenience we denote the numerical factors in these expressions by $A$ and $B$, respectively. Notice that $B$, the factor that appears in the Ricci tensor on $\Sigma_\ts{B}$, is strictly positive. The factor $A$, on the other hand, may be positive, negative, or zero, depending on the details of the theory.

Taken together, the fall-off conditions \eqref{metricBC}-\eqref{fluxBC} and the constraints \eqref{DilatonConstraint}-\eqref{Ricci1} on the boundary data are the ALD boundary conditions. There are two different motivations for this particular choice of boundary conditions. First, they are relevant for various supergravity theories. For example, the near horizon description of Neveu-Schwarz and Dirichlet 5-branes in type II supergravity take this form \cite{Narayan:2001dr, Parnachev:2005hh}. Second, they represent an interesting generalization of gravity with asymptotically flat (AF) boundary conditions \cite{Beig:1980be, Beig:1982xx, Ashtekar:1990gc, Ashtekar:1991vb, Mann:2005yr, Mann:2008ay}. This can be seen in the fall-off conditions for the metric, which are the same in both cases. Notice, though, that the constraints on the metric are very different in the two theories. For AF gravity the constraint on the metric $h_{ab}^\ms{(0)}$ at spatial infinity is
\begin{gather}\label{AFgravity}
\RR_{ab}^\ms{(0)} = (d-1) \, h_{ab}^{(0)} ~. 
\end{gather}
The analogous conditions \eqref{Ricci2} and \eqref{Ricci1} for ALD gravity are deformed due to the non-vanishing field strength at spatial infinity.

\subsection{Spacetime Cut-Offs and Regulated Integrals}
\label{sec:BndyDef}

The integrals that appear in the action \eqref{bulkaction} are not defined for fields that satisfy the fall-off conditions \eqref{metricBC}-\eqref{fluxBC}. For example, the bulk integral contains terms that diverge when the upper limit of the integral over $\rho$ is taken to infinity. Before we can verify the claim that this action is not suitable for spacetimes with non-compact spatial slices, we need to establish a systematic method for regulating these integrals.

We follow the same approach that was used in \cite{Mann:2005yr}. Let $\MM_\ms{\Omega} \subset \MM$ be a compact region with boundary $\dM_\ms{\Omega}$. The metric $g$ on $\MM$ induces a metric $h_\ms{\Omega} = g|_{\dM_\ms{\Omega}}$ on $\dM_\ms{\Omega}$. Now consider a one-parameter family of these regions that converges to $\MM$ as $\Omega \to \infty$; i.e., $\MM_\ms{\Omega'}\subset \MM_\ms{\Omega}$ for $\Omega > \Omega'$ and $\cup_\ms{\Omega} \MM_\ms{\Omega} = \MM$. For finite values of $\Omega$ this defines a method of cutting-off the spacetime. Integrals like the ones appearing in \eqref{bulkaction} are regulated by evaluating them on the compact region $(\MM_\ms{\Omega},g)$ or its boundary $(\dM_\ms{\Omega}, h_\ms{\Omega})$. The resulting functionals are perfectly well-defined for any finite value of the cut-off $\Omega$. Of course, the regulated integrals contain terms that diverge as $\Omega \to \infty$, and the cut-off cannot be removed until those terms have been addressed.

There are different families of cut-offs that can be used to regulate the action, with a family defined as a collection of cut-off prescriptions that all give the same results when the cut-off is removed. Therefore, the family of cut-offs used to regulate the action represents an additional structure that enters into the definition of the theory. Given the ALD boundary conditions for the theory, it is natural to focus on cut-offs that take the form $\Omega = \rho + \OO(\rho^{\,0})$\,\footnote{This is the hyperbolic cut-off for AF gravity discussed, along with other families of cut-offs, in \cite{Mann:2005yr}.}. With this choice the boundary of the compact region $\MM_\ms{\rho}$ is the constant $\rho$ surface $\Sigma$. Throughout the rest of this paper it is always assumed that integrals and functionals are regulated in this manner. When a functional is written explicitly in terms of integrals over $\MM$ and/or $\dM$, this means that the process of regulating the functional, evaluating it, and then removing the cut-off gives a finite result. If this is not the case, the functional will be expressed in terms of integrals over $\MM_\ms{\rho}$ and $\Sigma$.

\subsection{Variation of the Action}
\label{sec:ActionVariation}

Using the regulator defined in the last section, we can show that the action \eqref{bulkaction} is not stationary with respect to variations of the fields that preserve the ALD boundary conditions. The restriction of these variations to the regulating boundary $\Sigma$ take the form
\begin{align} \label{hvariation}
\delta h_{ab} = & \dub \rho^{\,4-d}\,\delta h_{ab}^\ms{(1)} + \rho^{\,3-d}\,\delta h_{ab}^\ms{(2)} \\ \label{phivariation}
\delta \varphi = & \dub \rho^{\,2-d}\,\delta \phi^\ms{(1)} + \rho^{\,1-d}\,\delta \phi^\ms{(2)} \\ \label{Avariation}
\delta \abig_{a_1 \ldots a_{p-1}} = & \dub \rho^{\,2-d}\,\delta \abig_{a_1 \ldots a_{p-1}}^\ms{(1)} +
    \rho^{\,1-d}\,\delta \abig_{a_1 \ldots a_{p-1}}^\ms{(2)} ~.
\end{align}
The scalars and tensors on the right-hand-side of these equations are arbitrary smooth fields on $\Sigma$ with the appropriate symmetries. The change in the action, evaluated on a solution of the equations of motion, is
\begin{align}\label{OnShellDeltaI}
 \delta I \big|_\ts{EOM} = &   \int_{\Sigma} \nts d^{\,d} x \, \sqrt{h} \, \left( \pi^{ab} \, \delta h_{ab}
 + \pi_{\phi} \, \delta \varphi + \pi^{a_1 \ldots a_{p-1}} \, \delta \abig_{a_1 \ldots a_{p-1}} \right) ~.
\end{align}
For the action to be stationary these surface terms must vanish as $\rho \to \infty$ for arbitrary field variations of the form \eqref{hvariation}-\eqref{Avariation}. To evaluate the surface terms we need the following expressions for the momenta
\begin{align}\label{hmomentum}
\pi^{ab} = & \dub h^{ab}\,K - K^{ab} \\ \label{phimomentum}
\pi_{\phi} = & \dub - \frac{8}{d-1}\,n^{a}\,\partial_{a} \phi \\ \label{Amomentum}
\pi^{a_{1}\ldots a_{p-1}} = & \dub - \frac{1}{2\,(p-1)!}\,n_{b}\,F^{b\,a_1 \ldots b_{p-1}} ~.
\end{align}
The boundary conditions on the $(p-1)$-form insure that the momentum \eqref{Amomentum} falls off very rapidly, so that terms in \eqref{OnShellDeltaI} proportional to $\delta \abig$ vanish as $\rho \to \infty$. This is not the case for the metric and dilaton. Their variation changes the action by
\begin{gather}\label{ActionVariationNoCT}
 \delta I \big|_\ts{EOM} = \int_{\Sigma} \nts d^{\,d}x \sqrt{h^\ms{(0)}} \, \left[
    (d-1)\,h^{ab}_\ms{(0)}\,\left( \rho \, \delta h_{ab}^\ms{(1)} + \delta h_{ab}^\ms{(2)} \right)
    - \frac{8\,(p-1)}{\alpha\,(d-1)}\,\left( \rho \, \delta \phi^\ms{(1)} + \delta \phi^\ms{(2)} \right)\right] ~.
\end{gather}
This does not vanish, despite the fact that the variations $\delta h_{ab}$ and $\delta \varphi$ `vanish at the boundary'. We conclude that the desired equations of motion are not stationary points of the action \eqref{bulkaction}. 

The result \eqref{ActionVariationNoCT} contains terms linear in $\rho$, so that the cut-off cannot be removed in this calculation. One might try to modify the boundary conditions so that these terms do not appear~\footnote{For example, the boundary conditions used in \cite{Marolf:2007ys} do not permit such terms.}. But even in that case \eqref{ActionVariationNoCT} still contains finite terms that do not vanish. Additional `pruning' of the boundary conditions results in a trivial variational principle, with the allowed field variations falling off too rapidly to distinguish between interesting solutions of the equations of motion. Instead, our only option is to modify the action.

\section{$\hK$ Counterterms}
\label{sec:KHat}

The fall-off conditions on the metric \eqref{metricBC} are the same as for an asymptotically flat spacetime in hyperbolic coordinates. This suggests that the $\hK$ counterterm constructed in \cite{Mann:2005yr} may also be relevant for the theories we are considering. The authors of \cite{Marolf:2007ys} used this approach -- with modifications -- to define an action for fields in a background of NS 5-branes in 10 dimensional type II supergravity. In this section we generalize the constructions of \cite{Mann:2005yr, Marolf:2007ys, Mann:2008ay} to the ALD theories described in the last section.

\subsection{$\hK$ for AF Gravity}
\label{sec:PureGravityKhat}

Let us briefly review the $\hK$ counterterm for gravity with AF boundary conditions. It was shown in \cite{Mann:2005yr, Mann:2006bd, Mann:2008ay} that solutions of the vacuum Einstein equations are stationary points of the action
\begin{gather}\label{puregravaction}
I = \int_{\MM}\bns d^{\,d+1}x \,\sqrt{g}\, R \, +\, 2 \int_{\dM} \bns \nts d^{\,d}x \,\sqrt{h}\,\left(K - \hK \right) ~.
\end{gather}
The term $\hK$ in the boundary integral is the trace of a symmetric tensor $\hK_{ab}$ defined (implicitly) by the equation
\begin{equation}\label{flatgravKhat}
 \hK \, \hK_{ab} - \hK_{a}{}^{c}\,\hK_{b c} = \RR_{ab} ~,
\end{equation}
with $\RR_{ab}$ the intrinsic Ricci tensor on $(\dM,h)$. The tensor $\hK_{ab}$ can be thought of as the part of the extrinsic curvature that is fixed by the kinematics of AF gravity. Furthermore, $\hK_{ab}$ is defined for any AF space time, which ensures that \eqref{puregravaction} has a well-defined variational problem.

In order to adapt this construction to ALD gravity, we first need to understand the origin of the defining equation \eqref{flatgravKhat}. The starting point is the expression \eqref{ElectricWeyl1} for the electric part of the bulk Weyl tensor evaluated on $\Sigma$
\begin{equation}\label{EWeyl1}
  E_{ab} = \perp(R_{a c b d} \, n^{c} n^{d}) - \perp(S_{a b}) -
    h_{ab} \, S_{c d} \, n^{c} n^{d} ~,
\end{equation}
where $n^{a}$ is the outward-pointing unit normal vector, $\perp$ represents projection of indices along $\Sigma$ via the induced metric $h_{ab}$, and $S_{ab}$ is the Schouten tensor defined in \eqref{Schouten}. Applying both the Gauss-Codazzi equations and the Einstein equations gives
\begin{equation}\label{EWeyl2}
K\,K_{ab} - K_{a}{}^{c}\,K_{b c} = \RR_{ab} + E_{ab} ~.
\end{equation}
This result does not hold for an arbitrary AF spacetime -- it only applies to solutions of the Einstein equations. However, all AF spacetimes must satisfy the kinematic constraints obtained from the leading term in an asymptotic expansion of the Einstein equations near spatial infinity. Therefore \emph{all} AF spacetimes satisfy the leading term in the asymptotic expansion of \eqref{EWeyl2}
\begin{equation}\label{preDefiningEqn}
K^\ms{(0)}\,K_{ab}^\ms{(0)} - K_{a}^\ms{(0)}{}^{c}\,K_{bc}^\ms{(0)} = \RR_{ab}^\ms{(0)} ~,
\end{equation}
which depends only on geometric identities and the kinematic constraints of the theory. 
The relation between  \eqref{preDefiningEqn} and \eqref{flatgravKhat} is apparent, but it will be useful to introduce one additional bit of notation before proceeding. Let $\widehat{=}$ denote equality between quantities on $\Sigma$ up to terms that are sub-leading at large $\rho$. Using this notation, the equation \eqref{preDefiningEqn} is equivalent to
\begin{equation}\label{preDefiningEqn2}
K\,K_{ab} - K_{a}{}^{c}\,K_{bc} \dub \widehat{=} \dub \RR_{ab} ~.
\end{equation}
The symmetric tensor $\hK_{ab}$ is defined as the tensor that satisfies \eqref{preDefiningEqn2} when `$\widehat{=}$' is replaced by `$=$'. This is the origin of the definition \eqref{flatgravKhat}.

\subsection{$\hK$ for ALD Gravity}
\label{sec:DefiningKhat}

This procedure outlined in the previous section can be generalized to ALD gravity. Starting from the definition \eqref{EWeyl1} and applying the equations of motion, we obtain
\begin{equation}\label{EWeyl3}
K\,K_{ab} - K_{a}{}^{c} \, K_{bc} = E_{ab} + \RR_{ab} -\left(\frac{d-2}{d-1}\right)\,\perp(T_{ab}) - \frac{1}{d\,(d-1)}\,h_{ab}\,T^{c}{}_{c} + \frac{1}{d-1}\,h_{ab}\,n^{c}\,n^{d}\,T_{cd}~,
\end{equation}
where $T_{ab}$ is the right-hand side of equation \eqref{EinsteinEOM}. The various projections of $T_{ab}$ can be worked out using the results of appendix \ref{sec:Hyper}. As before, we begin by expressing the leading term in the asymptotic expansion of this equation in terms of the boundary data. Using the `$\widehat{=}$' notation, the terms in \eqref{EWeyl3} that are relevant at leading order are
\begin{align}\label{preDefiningEqn3} \nonumber
K\,K_{ab} - K_{a}{}^{c} \, K_{bc} \dub \widehat{=} \,\, & \,\, E_{ab} + \RR_{ab} + \frac{4}{d\,(d-1)}\,h_{ab}\,\left(n^{c} \partial_{c} \phi \right)^{2} - \frac{(d-2)}{2\,(d-1)\,(p-1)!}\,e^{2\,\alpha\,\phi}\perp\left(F_{a}{}^{\ldots}\,F_{b \ldots} \right) \\ & \dub \dub + \frac{\big(p\,(d-2)-(d-1)\big)}{2\,d\,(d-1)\,p!}\,h_{ab}\,e^{2\,\alpha\,\phi}\,F^{\,2} ~.
\end{align}
Now the right-hand side of this equation must be expressed in terms of intrinsic quantities on $\Sigma$. This is straightforward for the field strength terms -- the boundary conditions allow us to replace $F_{a_1 \ldots a_p}$ with $\FF_{a_1 \ldots a_p}$, its projection along $\Sigma$. The normal derivative of $\phi$ can be expressed, up to sub-leading terms, as a simple function of $\phi$ on $\Sigma$
\begin{equation}
n^{c} \partial_{c} \phi \dub \widehat{=} \dub \frac{(p-1)}{\alpha}\,\beta\,e^{-\frac{\alpha}{p-1}\,\phi} ~.
\end{equation}
Likewise, the result \eqref{E0} gives the leading behavior of the electric part of the Weyl tensor as 
\begin{equation}
E_{ab} \dub \widehat{=} \dub - \frac{1}{d-1}\,\left(\RR_{ab} - \frac{1}{d}\,h_{ab}\,\RR \right) ~.
\end{equation}
Notice that this vanishes for the AF boundary conditions \eqref{AFgravity}, but not in the case of ALD gravity.

At this point \eqref{preDefiningEqn3} could be rewritten to furnish a defining equation for $\hK_{ab}$. However, in order to make a direct connection with the defining equation \eqref{flatgravKhat} for AF gravity, it is useful to express the leading part of $E_{ab}$ as a function of the dilaton and field strength on $\Sigma$. Using the various kinematic constraints on the fields, we obtain
\begin{equation}
E_{ab} \dub \widehat{=} \dub - \frac{1}{2\,(d-1)\,(p-1)!}\,e^{2\,\alpha\,\phi}\,\left( \FF_{a}{}^{\ldots}\,\FF_{b \ldots} -\frac{1}{d}\,h_{ab}\,\FF^{\,2} \right)~.
\end{equation}
Substituting this result in \eqref{preDefiningEqn3} now gives a defining equation for $\hK_{ab}$ that resembles the AF definition, with additional terms involving the dilaton and field strength
\begin{align}\label{DefiningEquation}
\hK\,\hK_{ab} - \hK_{a}{}^{c}\,\hK_{bc} = & \dub \RR_{ab} + \frac{4\,(p-1)^{2}\,\beta^{\,2}}{d\,(d-1)\,\alpha^{2}}\,h_{ab}\,e^{-\frac{2\,\alpha}{p-1}\,\phi} \\ \nonumber & - \frac{1}{2\,(p-1)!}\,e^{2\,\alpha\,\phi}\,\FF_{a}{}^{\ldots}\,\FF_{b \ldots} + \frac{(p-1)}{2\,d\,p!}\,h_{ab}\,e^{2\,\alpha\,\phi}\,\FF^{2} ~.
\end{align}
This equation can be solved order-by-order in the asymptotic expansion for any set of fields satisfying the ALD boundary conditions\,\footnote{In $3+1$ dimensions with AF boundary conditions it is possible to obtain $\hK_{ab}$ explicitly \cite{Visser:2008gx}.}. The result is an asymptotic expansion for $\hK_{ab}$ of the form
\begin{equation}\label{KhatExpansion}
  \hK_{ab} = \rho \,\hK_{ab}^\ms{(0)} + \rho^{\,3-d}\,\hK_{ab}^\ms{(1)} + \rho^{\,2-d}\,\hK_{ab}^\ms{(2)} + \ldots ~.
\end{equation}
The equation for the leading term is straightforward and gives
\begin{equation}\label{LeadingKhat}
  \hK_{ab}^\ms{(0)} = h_{ab}^\ms{(0)} ~,
\end{equation}
which shows that $\hK_{ab}$ and $K_{ab}$ agree at leading order\,\footnote{The equation \eqref{DefiningEquation} admits two solutions with opposite signs. We have chosen the solution whose sign agrees with that of the leading term in the extrinsic curvature.}. Expressions for the sub-leading terms in \eqref{KhatExpansion} are more complicated, and will be given as needed later in the paper.

An important caveat regarding the procedure described above is that it does not give a unique definition for $\hK_{ab}$\,! The boundary data for the different fields are related by the kinematic constraints in the boundary conditions, and these relations can be used to rewrite the right-hand side of \eqref{DefiningEquation}. For example, the boundary conditions imply the following asymptotic relations
\begin{align}\label{FManipulations}
  \frac{1}{2\,p!}\,e^{2\,\alpha\,\phi}\,\FF^{\,2} \dub \widehat{=} & \dub \frac{4\,(p-1)\,\beta^{\,2}}{\alpha^{2}}\,e^{-\frac{2\,\alpha}{p-1}\,\phi} \\ \label{RManipulations}
  \RR \dub \widehat{=} & \dub \RR^\ms{(0)} \, \beta^{\,2}\,e^{-\frac{2\,\alpha}{p-1}\,\phi}~.
\end{align}
These sorts of replacements will not change the value of $\hK_{ab}$ at leading order -- that is fixed by the boundary conditions -- but they \emph{will} change the functional dependence of $\hK_{ab}$ on the different fields. In other words, one obtains inequivalent definitions of $\hK_{ab}$. The obvious criteria for choosing one of these definitions is whether the corresponding counterterm, when added to the action \eqref{bulkaction}, leads to a well-defined variational principle. As we show in the next section, there is no definition that manages to do this.

\subsection{Variation of the Action}

We now investigate the variational properties of the action \eqref{bulkaction} with the additional surface term
\begin{equation}\label{hKterm}
I_{\hK} = -2\int_{\Sigma}\nts d^{\,d}x \,\sqrt{h}\,\hK ~.
\end{equation}
Specifically, we want to determine whether or not the equations of motion \eqref{EinsteinEOM}-\eqref{MaxwellEOM} are stationary points of the new action with respect to small changes in the fields that preserve the boundary conditions. The result \eqref{ActionVariationNoCT} showed that this was not the case for the action \eqref{bulkaction} on its own.

The change in \eqref{hKterm} due to a small variation of the fields is
\begin{equation}\label{deltaLKhat}
\delta I_{\hK} = -2\int_{\Sigma}\nts d^{\,d}x \,\sqrt{h}\,\left[\left(\frac{1}{2}\,\hK\,h^{ab} - \hK^{ab} \right)\,\delta h_{ab} + h^{ab}\,\delta \hK_{ab}\right] ~.
\end{equation}
The response $\delta\hK_{ab}$ is obtained by varying the defining equation \eqref{DefiningEquation}, and solving order-by-order in the asymptotic expansion. In fact, since it is the trace of $\delta\hK_{ab}$ that appears in \eqref{deltaLKhat}, it is sufficient to work with the trace of the defining equation
\begin{equation} \label{TraceDefEqn}
\hK^{2} - \hK^{ab}\,\hK_{ab} = \RR + \frac{4\,(p-1)^{2}}{\alpha^{\,2}\,(d-1)}\,\beta^{\,2} \, e^{-\frac{2\,\alpha}{p-1}\,\phi} - \frac{1}{2\,p!}\,e^{2\,\alpha\,\phi}\,\FF^{\,2} ~.
\end{equation}
Varying both sides of this equation and rearranging terms, we arrive at
\begin{align}\label{VarTraceDefEqn} \nonumber
2\,\left(\hK\,h^{ab} - \hK^{ab} \right)\,\delta \hK_{ab} = & \,\, \left(2\,\hK\,\hK^{ab} - 2\,\hK^{a}{}_{c}\,\hK^{bc} - \RR^{ab} + \frac{1}{2\,(p-1)!}\,e^{2\,\alpha\,\phi}\,\FF^{a}{}_{\ldots} \, \FF^{b \ldots} \right)\,\delta h_{ab} \\
 & \dub - \left(\frac{8\,(p-1)\,\beta^{\,2}}{\alpha\,(d-1)}\,e^{-\frac{2\,\alpha}{p-1}\,\phi} + \frac{\alpha}{p!}\,e^{2\,\alpha\,\phi}\,\FF^{\,2} \right) \,\delta \varphi + h^{ab}\,\delta \RR_{ab} \\ \nonumber
 & \dub - \frac{1}{(p-1)!}\,e^{2\,\alpha\,\phi}\,\FF^{a_{1} \ldots a_{p}}\,\DD_{a_{1}} \delta \abig_{a_{2}\ldots a_{p}} ~.
\end{align}
Only the leading term in this equation is relevant for determining \eqref{deltaLKhat}; the sub-leading terms fall off too rapidly to contribute. Furthermore, total derivatives in \eqref{VarTraceDefEqn} can be ignored because the term we are calculating is integrated over $\Sigma$\,\footnote{Terms in \eqref{VarTraceDefEqn} that contribute total derivatives at leading order also produce sub-leading terms that are \emph{not} total derivatives. These terms must be accounted for when solving for $\delta \hK_{ab}$ beyond leading order.}. Using the field variations \eqref{hvariation} - \eqref{Avariation}, the leading terms in \eqref{VarTraceDefEqn} are
\begin{align}\label{TraceKhatVariation}
h^{ab}_\ms{(0)} \, \delta \hK_{ab} = & \,\, \left( h^{ab}_{(0)} - \frac{1}{2\,(d-1)}\,\RR^{ab}_\ms{(0)} + \frac{2\,(p-1)}{\alpha^{\,2}\,(d-1)}\,\delta^{a}{}_{\alpha}\,\delta^{b}{}_{\beta}\,h^{\alpha\beta}_\ms{(0)} \right)\,\left(\rho^{\,3-d}\,\delta h_{ab}^\ms{(1)} + \rho^{\,2-d}\,\delta h_{ab}^\ms{(2)} \right) \\ \nonumber
& \dub - \frac{4\,d\,(p-1)}{\alpha\,(d-1)^{2}}\,\left( \rho^{\,3-d}\,\delta \phi^\ms{(1)} + \rho^{\,2-d}\,\delta \phi^\ms{(2)}\right) + \ldots ~.
\end{align}
The `$\ldots$' refer to terms that are either sub-leading or total derivatives. In particular, the terms involving the variation of the $p-1$ form are total derivatives. The result \eqref{TraceKhatVariation} applies for any field configuration that obeys the ALD boundary conditions, because it depends only on the defining equation and the constraints on the boundary data.

The new contribution to the variation of the action can be evaluated using \eqref{LeadingKhat} and \eqref{TraceKhatVariation}. Combining these with \eqref{ActionVariationNoCT} gives
\begin{align}\label{NewActionVariation}
\delta ( I + I_{\hK} ) = \int_{\Sigma} \nts d^{\,d}x \,\sqrt{h^\ms{(0)}}\,& \left[-\frac{4\,(p-1)^{2}}{\alpha^{\,2}\,(d-1)^{2}}\,h^{ab}_\ms{(0)}\, \left(\rho\,\delta h_{ab}^\ms{(1)} + \delta h_{ab}^\ms{(2)} \right) \right. \\ \nonumber
 & \left. \dub \dub + \frac{8\,(p-1)}{\alpha\,(d-1)^{2}}\,\left(\rho\,\delta \phi^\ms{(1)} + \delta \phi^\ms{(2)} \right) \right] ~.
\end{align}
This is qualitatively no different than the result we were trying to fix! The presence of the $\hK$ boundary term in the action \eqref{puregravaction} ensures a well-defined variational principle for gravity with AF boundary conditions, but the analogous construction does not work for theories with ALD boundary conditions. Of course, in AF gravity a unique definition of $\hK_{ab}$ follows from a combination of geometric identities and boundary conditions. But in the present case the definition is somewhat arbitrary -- terms on the right-hand side of \eqref{DefiningEquation} can be rewritten using relations between the boundary data. Does the result \eqref{NewActionVariation} mean that the $\hK$ construction does not work for ALD gravity, or does it tell us that we have chosen the wrong defining equation for $\hK_{ab}$?

We can answer this question by parameterizing the possible defining equations for $\hK_{ab}$, and showing that there is no definition such that \eqref{NewActionVariation} vanishes. First, note that the result \eqref{NewActionVariation} depends only on the trace of the defining equation. The trace belongs to a two-parameter family of equations, related to \eqref{TraceDefEqn} via \eqref{FManipulations} and \eqref{RManipulations}
\begin{align}
\hK^{2} - \hK^{ab}\,\hK_{ab} = & \dub (1+x)\,\RR - \frac{(1+y)}{2\,p!}\,e^{2\,\alpha\,\phi}\,\FF^{\,2} \\ \nonumber
& \dub + \left(\frac{4\,(p-1)^{2}}{\alpha^{\,2}\,(d-1)} - x\,\RR^\ms{(0)} + y\,\frac{4\,(p-1)}{\alpha^{\,2}}\right)\,\beta^{\,2}\,e^{-\frac{2\,\alpha}{p-1}\,\phi}
\end{align}
where $x$ and $y$ are arbitrary real numbers. Repeating the analysis of this section shows that there are no values of $x$ and $y$ for which $\delta I$ vanishes on-shell. This seems odd at first; there are two terms in \eqref{NewActionVariation} that need to be canceled, and two free parameters ($x$ and $y$) that can be adjusted. The problem becomes apparent when one chooses $x$ so that the $\delta \phi$ terms in \eqref{NewActionVariation} vanish. The variation of the action now contains terms proportional to $\delta h_{ab}$, with a coefficient that depends on the remaining free parameter $y$. But it also contains new terms proportional to $\delta h_{\alpha\beta}$, which were not present when \eqref{TraceDefEqn} was taken as the defining equation. There is no choice of $y$ that cancels the coefficients of both sets of terms. Therefore, it is not possible to obtain $\delta I = 0$ using only the $\hK$ counterterm. This same problem was encountered in \cite{Marolf:2007ys}, where it was addressed by adding additional surface terms to the action.

\subsection{Additional Boundary Terms}

Following \cite{Marolf:2007ys} we introduce three new surface terms, in addition to \eqref{hKterm}. These terms are functionals of quantities intrinsic to the boundary and take the form
\begin{align}\label{NewBoundaryTerms}
I_\ts{CT} = \int_{\Sigma} \nts d^{\,d}x\,\sqrt{h}\,\left(c_{_0} \, e^{-\frac{\alpha}{p-1}\,\phi} + c_{_1}\,e^{\frac{\alpha}{p-1}\,\phi}\,\RR + c_{_2}\,e^{\frac{2p-1}{p-1}\,\alpha\,\phi}\,\FF^{\,2} \right) ~.
\end{align}
The constants $c_{i}$ are determined by requiring that the full action satisfy $\delta I = 0$ on-shell. We will discuss a similar calculation in much more detail in the next section, so for now we simply note that for most values of $\alpha$ the condition $\delta I = 0$ requires
\begin{equation}\label{deltaIconditions}
c_{0} = \frac{8\,(p-1)^{2}\,\beta}{\alpha^{\,2}\,(d-1)^{2}} \quad \quad c_{1} = c_{2} = 0 ~.
\end{equation}
There is also a special case, $\alpha = \pm 2(p-1)/(d-1)$, where the equations that determine the $c_i$ become degenerate and admit a one-parameter family of solutions. In that case the $c_i$ are given by
\begin{equation}\label{deltaIconditions2}
c_{0} = 2\,\beta - (d-1)^{2}\,\beta^{\,2}\,c_{1} \quad \quad  c_{2} = -\frac{1}{2\,p!}\,c_{1} \end{equation}
with $c_1$ arbitrary. The special values of $\alpha$ corresponds to the coupling between the dilaton and the NS-NS\,(-) or R-R\,(+) 3-form field strengths in ten dimensional type II supergravity.

The authors of \cite{Marolf:2007ys} considered the NS-NS sector of type II supergravity in 10 dimensions.  However, instead of adding the third boundary term in \eqref{NewBoundaryTerms} they included a term proportional to  $\RR\,\FF^{\,2}$. This raises an important question: what if we did not limit ourselves to the three boundary terms in \eqref{NewBoundaryTerms}? For instance, one might add the following boundary terms to the action
\begin{align}\label{NewBoundaryTerms2}
I_\ts{CT} = \int_{\Sigma} \nts d^{\,d}x\,\sqrt{h}\,\left(c_{_0} \, e^{-\frac{\alpha}{p-1}\,\phi} + c_{_1}\,e^{\frac{\alpha}{p-1}\,\phi}\,\RR + c_{_2}\,e^{\frac{2p-1}{p-1}\,\alpha\,\phi}\,\FF^{\,2} + c_{4}\,e^{\frac{3\,\alpha}{p-1}\,\phi}\,\RR^{2} + \ldots  \right)
\end{align}
For generic values of $\alpha$ this leads to a one-parameter family of actions such that $\delta I = 0$, and for the special value of $\alpha$ mentioned above a two-parameter family is obtained. This is qualitatively different than what one finds in asymptotically anti-de Sitter spacetimes, where a higher derivative boundary term like $\RR^{2}$ always falls off faster than a term like $\RR$. In a theory with ALD boundary conditions, dressing the higher derivative boundary terms with suitable functions of the dilaton ensures that all of the counterterms contribute at the same orders in the asymptotic expansion. This means that one may add an arbitrary number of boundary terms to obtain a family of actions, all of which satisfy $\delta I = 0$, with a corresponding number of free parameters! We will return to this issue later. For now, we focus on the boundary terms \eqref{NewBoundaryTerms}.

\subsection{Finiteness of the On-Shell Action}

The conditions \eqref{deltaIconditions} or \eqref{deltaIconditions2} also imply that the on-shell action is finite in the $\rho \to \infty$ limit. The equations of motion can be used to rewrite the bulk terms in the action as a total derivative, so that the on-shell action is given by
\begin{align}\label{OnShellAction}
I \big|_\ts{EOM} = \int_{\dM} \bns d^{\,d}x \, \sqrt{h}\,& \bigg[  \raisebox{16pt}{} - \frac{8\, (p-1)}{\alpha\,(d-1)^{2}}\,n^{a}\,\partial_{a}\phi + 2\,K
 - 2\,\hK + c_{_0} \, e^{-\frac{2\,\alpha}{p-1}\,\phi} \\ \nonumber
 &  + c_{_1}\,e^{-\frac{2\,\alpha}{p-1}\,\phi}\,\RR + c_{_2}\,e^{-\frac{2p-1}{p-1}\,\alpha\,\phi}\,\FF^{\,2} \raisebox{16pt}{} \bigg] + \ldots ~,
\end{align}
where `$\ldots$' represents a contribution from the total derivative evaluated at the lower limit of the bulk integral. These terms can be ignored, since they do not depend on the cut-off $\rho$ and we are only interested in establishing finiteness at this point. Using the asymptotic expansions of the various boundary terms, we find that the on-shell action contains a leading term of order $\rho^{d-1}$ and a sub-leading term proportional to $\rho$. The remaining terms are finite, or vanish as the cut-off is removed. It is straightforward to show that the leading term in the expansion vanishes as a result of the boundary conditions and \eqref{LeadingKhat}. The order $\rho$ term needs to cancel as well, and to show this we need the first sub-leading term in the asymptotic expansion of $\hK$. The asymptotic expansion of $\hK$ is
\begin{equation}
 \hK= d\,\rho^{-1} + \rho^{\,1-d}\,\left(h^{ab}_\ms{(0)}\hK_{ab}^\ms{(1)}- h^\ms{(1)} \right) + \OO(\rho^{-d}) ~,
\end{equation}
where we have used the result \eqref{LeadingKhat} for the leading term, and terms of $\OO(\rho^{-d})$ give finite contributions to the action. Solving \eqref{TraceDefEqn}, we find that the order $\rho^{1-d}$ term in $\hK$ is
\begin{align}
\hK^{(1)} -  h^{(1)} = \frac{1}{2\,(d-1)}\,& \left(\RR^\ms{(1)} - h^{ab}_\ms{(1)}\,\RR_{ab}^\ms{(0)}
-\frac{8\,(p-1)}{\alpha\,(d-1)}\,\phi^\ms{(1)} \right. \\ \nonumber
 & \left. \quad - \frac{4\,(p-1)}{\alpha^{\,2}}\,\left(Y^\ms{(1)} - h^\ms{(1)}_{s} + 2\,\alpha\,\phi^\ms{(1)} \right) \right) ~
\end{align}
where $h^\ms{(1)}_{s} = h^{\alpha\beta}_\ms{(0)}\,h_{\alpha\beta}^\ms{(1)}$, and $Y^\ms{(1)}$ is defined in equation \eqref{YDef}. Using this result in \eqref{OnShellAction} shows that the order $\rho$ term in the on-shell action vanishes. For generic values of $\alpha$, the conditions \eqref{deltaIconditions} and the equations of motion in appendix \ref{sec:EOMandBianchi} imply that this term vanishes identically. In the special cases $\alpha = \pm2(p-1)/(d-1)$, terms at this order vanish up to total derivatives. In either case, the lesson is that finiteness of the on-shell action follows from the requirement that the action admits a well-defined variational principle.

\section{Explicit Counterterms}
\label{sec:OC}

The $\hK$ construction is of interest because it makes contact with a procedure that works for gravity with AF boundary conditions. But this approach is not sufficient if the theory contains additional fields with support at spatial infinity. In that case new surface terms must be included in the action. These terms contribute to the action and its variation at the same orders in the asymptotic expansion as the $\hK$ term; it stands to reason that one might be able to define a viable action using \emph{only} terms of the form \eqref{NewBoundaryTerms}.

In this section we drop the $\hK$ counterterm and construct actions using boundary counterterms that are defined explicitly as local functionals of the fields on the boundary. The results are reminiscent of the counterterms used in the AdS/CFT correspondence \cite{Henningson:1998gx, Balasubramanian:1999re, Emparan:1999pm, Bianchi:2001kw, Martelli:2002sp, Batrachenko:2004fd, Papadimitriou:2005ii}, but there are also significant differences. In AdS/CFT a well-defined variational principle identifies a finite set of counterterms \cite{Papadimitriou:2005ii}. For ALD theories we encounter families of actions that admit a well-defined variational problem but depend on an arbitrary number of free parameters. Surprisingly, we find that the on-shell action and the conserved charges do not depend on the choice of action. This is reassuring, because even in cases where there are compelling reasons to limit ourselves to a small number of possible counterterms, we still find interesting examples that admit a one-parameter family of actions.

\subsection{The Action}
\label{sec:ECAction}

Consider a counterterm action that takes the following form
\begin{equation}\label{CT}
  I_\ts{CT} = \int_{\Sigma} \nts d^{\,d}x \, \sqrt{h}\,\bigg(U(\phi) + C(\phi) \, \RR + Z(\phi)\,\FF^{a_1 \ldots a_p} \FF_{a_1 \ldots a_p} + \ldots \bigg) ~.
\end{equation}
This can be thought of as a derivative expansion on $\Sigma$, neglecting terms that vanish due to the boundary conditions. For example, a term of the form $M(\phi) (D \phi)^2$ will not contribute to the action or its first variation. At this point we restrict our attention to terms containing no more than two (boundary) derivatives of the fields on $\Sigma$. This is motivated by our interest in various supergravity theories obtained as the low-energy limit of string theory: we do not wish to include surface terms corresponding to higher orders in the $\alpha'$ expansion than the terms that appear in the bulk part of the action.

Including the counterterms \eqref{CT}, the on-shell variation of the action is
\begin{align}\label{1stVariation}
 \delta I \big|_\ts{EOM} = \int_{\dM} \bns d^{\,d}x \,\sqrt{h}
    \, & \bigg[ \left( \pi^{ab} + P^{ab} \right)\,\delta h_{ab} + \left( \pi_{\phi} + P_{\phi} \right) \delta \phi \\ \nonumber
    & \quad + \left(\pi^{a_1 \ldots a_{p-1}} + P^{a_1 \ldots a_{p-1}} \right)
    \,\delta\abig_{a_1 \ldots a_{p-1}} \bigg]
\end{align}
where $\pi^{ab}$, $\pi_{\phi}$, and $\pi^{a_1 \ldots a_{p-1}}$ are the momenta given in \eqref{hmomentum}-\eqref{Amomentum}. The terms $P^{ab}$, $P_{\phi}$, and $P^{a_1 \ldots a_{p-1}}$ are functional derivatives of \eqref{CT}
\begin{align} \label{CTMomentum1}
P^{ab}  = \dub & \frac{1}{2}\,h^{ab}\,U - C \left(\RR^{ab} - \frac{1}{2}\,h^{ab}\,\RR \right) + D^{a} D^{b} C - h^{ab}\,D^{2}C\\  \nonumber
& \dub + Z \, \left( \frac{1}{2}\,h^{ab}\,\FF^{\,2} - p\,\FF^{a}{}_{\ldots}\,\FF^{b \ldots} \right) \\
 \label{CTMomentum2}
P_{\phi} = \dub & \frac{\partial U}{\partial \phi} + \frac{\partial C}{\partial \phi} \, \RR + \frac{\partial Z}{\partial \phi} \, \FF^{\,2}  \raisebox{12pt}{}\\ \label{CTMomentum3}
P^{a_1 \ldots a_{p-1}} = \dub & -2\,p\,D_{b}\big(Z \,\FF^{b a_{1} \ldots a_{p-1}} \big)\raisebox{12pt}{}
\end{align}
The functions $U$, $C$, and $Z$ are determined by the requirement that \eqref{1stVariation} vanishes for field variations that preserve the boundary conditions. Based on the results of the previous section, these functions are expected to take the form
\begin{equation}
  U(\phi) = c_0 \, e^{- \frac{\alpha}{p-1} \, \phi} \quad\quad C(\phi) = c_1 \, e^{
    \frac{\alpha}{p-1} \, \phi}
    \quad\quad Z(\phi) = c_2 \, e^{\frac{2p-1}{p-1} \, \alpha\, \phi} ~.
\end{equation}
The constants $c_i$ are determined by canceling the appropriate terms in \eqref{1stVariation}. It is interesting to note that the dilaton dependence in these expressions means that the terms in \eqref{CT} correspond to an expansion in powers of the inverse metric in string frame.

As before, the boundary conditions on the gauge field imply that the terms proportional to $\delta \abig$ fall off very rapidly and do not contribute to \eqref{1stVariation}. This leaves the terms proportional to $\delta h_{ab}$ and $\delta \phi$. The leading behavior of the coefficients of these terms are
\begin{align}\label{metriccancel}
\pi^{ab} + P^{ab} = \rho^{-3} & \left(d-1 + \frac{c_0}{2\,\beta} - c_1\,\beta\,A + \frac{1}{2}\,c_1\,\beta\,\big(A \,(d-p) + B\,p\big) +c_2\,\beta\,p!\,(B-A) \right)\,h^{ab}_\ms{(0)} \\ \nonumber
 & \quad + \beta \, \rho^{-3}\,\big((B-A)\,c_1 + c_2\,p!\,Q^{2}\,\beta^{\,2(p-1)} \big)\,\delta^{a}{}_{\alpha} \,\delta^{b}{}_{\beta}\,h^{\alpha\beta}_\ms{(0)} \\ \label{dilatoncancel}
\pi_{\phi} + P_{\phi} = \rho^{-1} & \frac{\alpha}{p-1}\,\bigg(-\frac{8\,(p-1)^2}{\alpha^{\,2}\,(d-1)} - \frac{c_0}{\beta} + c_1 \, \beta\, \big(A\,(d-p) + B\,p\big) \\ \nonumber 
  & \quad \quad \quad \dub  + 2\,c_2\,\beta\,p!\,(2p-1)\,(B-A)\bigg)
\end{align}
where $A$ and $B$ are the numerical factors appearing in the expressions \eqref{Ricci2} and \eqref{Ricci1} for the Ricci tensor on $\Sigma$.  The $\rho \to \infty$ limit of \eqref{1stVariation} vanishes if these expressions are set to zero. This yields three  equations, because the terms in \eqref{metriccancel} with support along $\Sigma_\ts{B}$ must vanish independently. These equations uniquely determine the $c_i$ for most values of $\alpha$
\begin{gather}\label{uniqueci}
 c_0 = \beta\,\left(-d + \frac{4\,(p-1)^2}{\alpha^{\,2}\,(d-1)^{2}} \right) \quad \quad c_1 = - \frac{1}{\beta\,(d-1)}
 \quad \quad c_2 = \frac{1}{2\,\beta\,(d-1)\,p!} ~.
\end{gather}
However, two of the conditions on the $c_i$ become linearly dependent when $\alpha$ takes the special values 
\begin{equation}\label{SpecialAlpha}
  \alpha = \pm \,2 \, \left(\frac{p-1}{d-1}\right) ~.
\end{equation}
In that case there is a one-parameter family of solutions 
\begin{gather}\label{1ParamFamily}
c_{0} = -2\,(d-1)\,\beta - c_1\,(d-1)^{2}\,\beta^{\,2} \quad \quad c_2 = - \frac{1}{2\,p!}\,c_1 ~.
\end{gather}
This result is relevant for type II supergravity in ten dimensions, as it corresponds to the coupling between the dilaton and the 3-form field strengths in the NS-NS and R-R sectors.

To summarize, we have found that adding the surface terms
\begin{align}\label{FinalCTs}
 I_\ts{CT} = \int_{\Sigma} \nts d^{\,d}x \,\sqrt{h}\,& \left[ \left(\frac{4}{\alpha^{2}}\,
   \left(\frac{p-1}{d-1} \right)^2 - d \right) \beta\,e^{-\frac{\alpha}{p-1}\,\phi}
   - \frac{1}{\beta\,(d-1)}\,e^{\frac{\alpha}{p-1}\,\phi} \left(\RR
   - \, \frac{1}{2\,p!} \, e^{2\,\alpha\,\phi}
  \, \FF^{\,2}\right)
  \right]
\end{align}
to the action \eqref{bulkaction} leads to a well-defined variational problem for ALD
boundary conditions. Notice that the $\RR$ and $\FF^{\,2}$ terms appear in \eqref{FinalCTs} with the same relative coefficients as the $R$ and $F^{2}$ term in the bulk part of the action, including the factor of $e^{2\,\alpha\,\phi}$. It is straightforward to show that the full action is finite on-shell. Using the same arguments as in the previous section, we can write the on-shell bulk term as a total derivative. The surface terms in the on-shell action are then given by
\begin{align}\label{OnShellSurface}
  I \big|_\ts{EOM} = \int_{\dM} \bns d^{\,d} x \,\sqrt{h} \, & \left[
    - \frac{8\,(p-1)}{\alpha\,(d-1)^2}\,n^{a}\, \partial_{a} \phi + 2\,K + \left(\frac{4}{\alpha^{2}}\,
   \left(\frac{p-1}{d-1} \right)^2 - d \right)\,\beta\,e^{-\frac{\alpha}{p-1}\,\phi}
   \right. \\ \nonumber
   & \dub \left. - \frac{1}{\beta\,(d-1)}\,e^{\frac{\alpha}{p-1}\,\phi}\,\RR + \frac{1}{2\,\beta\,p!\,(d-1)}
   \, e^{\frac{2p-1}{p-1}\,\alpha\,\phi} \, \FF^{\,2} \raisebox{20pt}{~}\right]  ~.
\end{align}
Of course, the on-shell action may also receive contributions from the total derivative evaluated at the lower limit of the bulk integral, but those terms are manifestly independent of the cut-off. Using the asymptotic expansions of the various fields in \eqref{OnShellSurface} gives
\begin{align}
I \big|_\ts{EOM} = \int_{\dM} \bns d^{\,d}x \, \sqrt{h^{_{(0)}}}\, \rho^{\,d-1} & \left[
    - \frac{8\,(p-1)^2}{\alpha^{2}\,(d-1)^{2}} + 2\,d + \frac{1}{\beta}\,c_0
    + \beta\,\RR^\ms{(0)}\,c_1 \right. \\ \nonumber
    & \dub \left. + \beta\,p!\,\frac{8\,(p-1)}{\alpha^{2}}\,c_2 +
    \OO(\rho^{1-d}) \,\right] ~.
\end{align}
The leading term vanishes when evaluated using the expressions for $\RR^\ms{(0)}$ and the
$c_i$, so that the action is finite as $\rho \to \infty$. Once again, finiteness of the on-shell action follows from the requirement that the action possess the proper variational properties.

\subsection{Higher Derivative Counterterms}
\label{sec:HDC}

 The counterterms used in the previous section contain no more than two derivatives of the fields. This seems appropriate for theories obtained as the low-energy limit of some quantum theory of gravity: in that case the boundary terms in the action should contain no more derivatives than the bulk terms. But relaxing this assumption reveals that there are many possible counterterms that lead to a well-defined variational principle for ALD boundary conditions. This is closely related to the ambiguity in the action when $\alpha$ takes the special value \eqref{SpecialAlpha}, so it is worth examining the issue more closely.

To illustrate how families of actions can be constructed, suppose that we remove the restriction on the number of derivatives that can appear in the counterterms and make the following ansatz for $I_\ts{CT}$
\begin{equation}\label{HDCT}
  I_\ts{CT} = \int_{\dM} \bns d^{\,d}x \, \sqrt{h}\,\left( b_0 \, e^{- \frac{\alpha}{p-1} \, \phi} + b_1 \, e^{\frac{\alpha}{p-1} \, \phi}\,\RR + b_2 \, e^{\frac{2p-1}{p-1} \, \alpha\,\phi}\,\FF^{a_1 \ldots a_p} \FF_{a_1 \ldots a_p} + b_3 \, e^{\frac{3\,\alpha}{p-1}\,\phi}
 \, \RR^{2} \right) ~.
\end{equation}
This differs from \eqref{CT} by an $\RR^{2}$ term, with a dilaton prefactor chosen to give the appropriate $\rho$ dependence. The coefficients are labeled $b_i$ to emphasize that this set of counterterms is distinct from the previous result \eqref{uniqueci}. The momenta $P_{\phi}$ and $P_{ab}$ obtained from this action are similar to \eqref{CTMomentum1} and \eqref{CTMomentum2}, with the $c_i$ replaced by $b_i$ and the following additional terms
\begin{align}
 \Delta P_{\phi} & = b_3 \, \frac{3\,\alpha}{p-1} \, e^{\frac{3\,\alpha}{p-1}\,\phi}
 \, \RR^{2} \\
 \Delta P^{ab}  & = b_3 \, e^{\frac{3\,\alpha}{p-1}\,\phi} \, \left( \frac{1}{2} \, h^{ab} \, \RR^{2} -
    2 \, \RR \, \RR^{ab} \right) + 2\,b_{3}\,D^{a}D^{b}\left(e^{\frac{3\,\alpha}{p-1}\,\phi}\,\RR\right)
    -2\,b_{3}\,h^{ab}\,D^{\,2} \left( e^{\frac{3\,\alpha}{p-1}\,\phi}\,\RR \right)~.
\end{align}
Requiring $\delta I|_\ts{EOM} = 0$ gives three equations for the four constants $b_i$. The one-parameter family of solutions can be expressed in terms of the previous solution \eqref{uniqueci} as
\begin{equation}\label{HDCTCoefficients}
b_0 = c_0 + b_3 \, \beta^{\,4}\,\RR_\ms{(0)}^{\,2} \quad \quad b_{1} = c_{1} - 2\,b_3\,\beta^{\,2}\,\RR_\ms{(0)} \quad \quad b_{2} = c_{2} ~,
\end{equation}
where $b_3$ an arbitrary real number. In the special case $\alpha = \pm 2(p-1)/(d-1)$ this procedure gives a two-parameter family of counterterms.

The full action with the new set of counterterms exhibits two important properties. First, despite the presence of the new surface term, the on-shell action remains finite for all values of the coefficient $b_3$. The relationships \eqref{HDCTCoefficients} ensure that a divergent contribution from the $\RR^2$ counterterm is canceled by the other surface terms. In other words, a well-defined variational principle may not determine a unique action, but it is sufficient to render the on-shell action finite. Second, the value of the on-shell action remains the same for all values of $b_3$. The shifts in the coefficients \eqref{HDCTCoefficients} and the new counterterm give additional finite contributions to the on-shell action, but these terms combine to form total derivatives or otherwise vanish by the equations of motion. Both of these results generalize as more counterterms are added to the action.  Thus, allowing counterterms with more than two derivatives leads to families of actions with a well-defined variational principle that depend on an arbitrary number of free parameters. However, all of these actions are finite and have the same on-shell value. 

When the parameter $\alpha$ takes the special value \eqref{SpecialAlpha}, restricting to two-derivative counterterms still results in a one-parameter family of actions. The results described above apply to this case as well: the action is finite, and its on-shell value is independent of the free parameter $c_1$ in \eqref{1ParamFamily}. In that case it is still possible that other considerations might identify a unique action. For instance, requiring the full action to exhibit a particular symmetry or duality might constrain the sorts of surface terms that can be introduced. This is known to be the case in other examples of holographic renormalization, such as dilaton gravity in two dimensions. A construction similar to the one described above \cite{Davis:2004xi, Grumiller:2007ju} yields a one-parameter family of counterterms, but only a particular value of the free parameter preserves Buscher duality \cite{Buscher:1987qj} in the full action.

\section{Linear Response Functions and Conserved Charges}
\label{sec:LRF}

In previous sections we were concerned with the properties of the action under field variations that preserve the ALD boundary conditions. Now we focus our attention on general variations of the fields. The resulting change in the action is characterized by a set of linear response functions related to the coefficients of the field variations in \eqref{1stVariation}.

\subsection{The Stress Tensor}
\label{sec:ST}

The linear response functions include a boundary stress tensor, obtained by varying the on-shell action with respect to the boundary metric
\begin{gather}\label{BST}
T^{ab} \defeq \left. \frac{2}{\sqrt{h}}\,\frac{\delta \, I}{\delta \, h_{ab}}\right|_\ts{EOM} = 2\,\left( \pi^{ab} + P^{ab} \right) ~.
\end{gather}
This differs from the definition used by Brown and York \cite{Brown:1992br, Brown:2000dz}, who obtain a quasi-local stress tensor via a Hamilton-Jacobi variation of the action\,\footnote{A variation within the space of classical field configurations due to a small change in the boundary data}. Their approach is not possible in the present case, because the boundary data for the metric is restricted by the kinematic constraints \eqref{Ricci2} and \eqref{Ricci1} in the ALD boundary conditions. Instead we adopt the construction \eqref{BST}, which was used in \cite{Mann:2005yr, Mann:2006bd, Mann:2008ay} and gives a sensible definition of conserved quantities. 

The asymptotic expansion of the boundary stress tensor is given by
\begin{gather}\label{StressTensorExpansion}
T_{ab} = \rho^{\,3-d}\,T_{ab}^\ms{(1)} + \rho^{\,2-d}\,T_{ab}^\ms{(2)} + \OO(\rho^{1-d}) ~.
\end{gather}
The various terms are evaluated using the asymptotic expansions of the tensors $\pi^{ab}$ and $P^{ab}$, and the equations of motion. For generic values of $\alpha$ we obtain
\begin{gather}\label{genericSTpot}
T_{ab}^\ms{(1)} =  \frac{\,(d-2)^2}{(d-1)}\,\gamma_{ab}^\ms{(1)} - \frac{2\,\alpha}{(p-1)(d-1)}\,\delta_{ab}^\ms{(1)}\\
T_{ab}^\ms{(2)} =  (d-1)\,\gamma_{ab}^\ms{(2)} - \frac{2\,\alpha}{(p-1)(d-1)} \, \delta_{ab}^\ms{(2)} ~.
\end{gather}
The symmetric tensors $\gamma_{ab}^\ms{(n)}$ and $\delta_{ab}^\ms{(n)}$ are defined as
\begin{gather} \label{gammadef}
 \gamma_{ab}^\ms{(n)} = h_{ab}^\ms{(n)} - h_{ab}^\ms{(0)}\,h^\ms{(n)} +
   \frac{8\,(p-1)}{\alpha\,(d-1)\,(d+n-3)}\,h_{ab}^\ms{(0)}\,\phi^\ms{(n)} \\ \label{deltadef}
 \delta_{ab}^\ms{(n)} = \DD_{a} \DD_{b} \phi^\ms{(n)} - h_{ab}^\ms{(0)}\,\DD^{2} \phi^\ms{(n)} -
   A\,h_{ab}^\ms{(0)}\,\phi^\ms{(n)} ~,
\end{gather}
where $\DD$ is the boundary covariant derivative compatible with the metric $h_{ab}^\ts{(0)}$, and $A$ is the numerical coefficient in the result \eqref{Ricci2} for the Ricci tensor on $\Sigma_\ts{A}$. For the special case $\alpha = \pm 2 \, (p-1)/(d-1)$, the components of the stress tensor are instead given by
\begin{align}\label{specialST1}
T_{ab}^\ms{(1)} = & \,\, (d-2)\,\gamma_{ab}^\ms{(1)} + c_1\,\beta\, (d-2)\,\gamma_{ab}^\ms{(1)} - \frac{4\,c_1 \, \beta}{(d-1)}\, \delta_{ab}^\ms{(1)} \\ \label{specialST2}
T_{ab}^\ms{(2)} = & \,\,(d-1)\,\gamma_{ab}^\ms{(2)} - \frac{4\,c_1 \, \beta}{(d-1)}\,\delta_{ab}^\ms{(2)} ~.
\end{align}
Despite the appearance of the free parameter $c_1$ in these expressions, we show in the next section that the conserved charges are independent of the choice of counterterms.

\subsection{Conserved Charges}
\label{sec:CC}

Consider a $d-1$ dimensional, spacelike surface $\CC \subset \Sigma$, with timelike unit normal vector $u^{a}$. We will assume that $u^{a}$ lies in $\Sigma_\ts{A}$, so that $u^{a}\,v_{a} = 0$ for any $v_{a}$ in $\Sigma_\ts{B}$. If $\xi^{a}$ is an asymptotic Killing vector that preserves the ALD boundary conditions, then the conserved charge associated with $\xi^{a}$ is given by the flux of the current $T_{ab}\,\xi^{b}$ across $\CC$
\begin{equation}\label{Charges}
  Q[\xi] = \int_{\,\CC} d^{\,d-1}x \,\sqrt{h_\ms{\CC}}\,u^{a}\,T_{ab}\,\xi^{b} ~.
\end{equation}
Conserved charges constructed in this manner necessarily generate the asymptotic symmetries of the theory \cite{HIM2}. It is straightforward to show that the $\xi^{a}$ that preserve the ALD boundary conditions are given by the Killing vectors of $h_{ab}^\ms{(0)}$, up to sub-leading terms that generate gauge transformations. The asymptotic expansion of these vectors is\,\footnote{In AF gravity, asymptotic boosts are associated with terms of order $\rho^{-1}$ in $\xi^{a}$. These terms are not present for ALD boundary conditions because they do not preserve the asymptotic form of the dilaton.}
\begin{gather}
\xi^{a}(\rho,x) = \xi^{a}_\ms{(0)}(x) + \OO(\rho^{2-d}) ~.
\end{gather}
Given this result, and the expansion \eqref{StressTensorExpansion} for the boundary stress tensor, only the leading terms in the asymptotic expansions of $h_\ms{\CC}$ and $u^{a}$ contribute to the conserved charges
\begin{equation}\label{QExpression}
  Q[\xi] = \int_{\CC} \nts d^{\,d-1}x \,\sqrt{h_\ms{\CC}^\ms{(0)}}\,\left( \rho\,u^{a}_\ms{(0)}\,T_{ab}^{(1)}\,\xi^{b}_\ms{(0)}  + u^{a}_\ms{(0)}\,T_{ab}^{(2)}\,\xi^{b}_\ms{(0)} + \OO(\rho^{-1}) \right) ~,
\end{equation}
where $h_{ab}^\ms{(0)}\,u^{a}_\ms{(0)}\,u^{b}_\ms{(0)} = -1$. Obviously, this is only well-defined if the first term in the integrand has a vanishing integral over $\CC$. Then the remaining term, which is finite as $\rho \to \infty$, gives the conserved charge associated with $\xi^{a}$. We will now show that this is indeed the case.

\vskip1em
\noindent{\bf Scalar and Tensor Potentials}
\vskip.5em

\noindent In order to prove that the conserved charges are well-defined and independent of any free parameters in the action, it is useful to first establish several results concerning potentials for symmetric tensors on $\Sigma$. Terms in the boundary stress tensor that admit such potentials do not contribute to the conserved charges. See \cite{Mann:2008ay} for a recent discussion in the context of AF gravity.

A tensor $\theta_{ab}$ is said to admit a scalar potential $\psi$ if it can be expressed in the form
\begin{gather}\label{ScalarPotential}
  \theta_{ab}[\psi] = \DD_{a} \DD_{b} \psi - h_{ab}^\ms{(0)}\,\DD^{2} \psi - A\,h_{ab}^\ms{(0)}\,\psi ~ ,
\end{gather}
Taking the covariant divergence we find that $\theta_{ab}$ is not, strictly speaking, conserved. However, its divergence is restricted to lie in the $p$-dimensional subspace $\Sigma_\ts{B} \subset \Sigma$
\begin{gather}
\DD^{b} \theta_{ab} = ( B - A ) \,\delta_{a}{}^{\alpha}\,\DD_{\alpha} \psi ~.
\end{gather}
As a result the tensor $\theta_{ab}$ is `effectively conserved' in the following sense: the current $\theta_{ab} \, \xi_\ms{(0)}^{b}$, where $\xi_\ms{(0)}^{b}$ is a Killing vector of the metric $h_{ab}^\ms{(0)}$, does not contribute to conserved charges. Such currents take the form
\begin{gather}\label{thetacurrent}
\theta_{ab} \, \xi_\ms{(0)}^{b} = \DD^{b} \left(2\,\xi^\ms{(0)}_{[b} \DD^{\vphantom{\ms{(0)}}}_{a]}\psi + \psi \, \DD^{\vphantom{\ms{(0)}}}_{[b} \,\xi_{a]}^\ms{(0)} \right) + (B-A)\,\delta_{a}{}^{\alpha}\,\xi_{\alpha}^\ms{(0)}\,\psi ~.
\end{gather}
This expression is obtained using properties of the Killing vector, which imply $\DD^{2}\xi_\ms{(0)}^{b} = -\RR^{b c}_\ms{(0)}\,\xi_{c}^\ms{(0)}$. Now consider the flux of the current across $\CC$, which is given by the integral of $u_\ms{(0)}^{a} \theta_{ab} \xi_\ms{(0)}^{b}$. The last term in \eqref{thetacurrent} does not contribute because $u_\ms{(0)}^{a}\,\delta_{a}{}^{\alpha} = 0$. The remaining term takes the form
\begin{gather}
u_\ms{(0)}^{a}\,\theta_{ab}\,\xi_\ms{(0)}^{b} = \DD^{b}_\ms{\CC} \left(2\,u_\ms{(0)}^{a}\,\xi^\ms{(0)}_{[b} \DD^{\vphantom{\ms{(0)}}}_{a]}\psi + u_\ms{(0)}^{a}\,\psi \, \DD^{\vphantom{\ms{(0)}}}_{[b} \, \xi^\ms{(0)}_{a]} \right)
\end{gather}
where $\DD_\ms{\CC}^{a}$ is the covariant derivative compatible with the induced metric on $\CC$. The integral vanishes, so a current of the form \eqref{thetacurrent} has vanishing flux across $\CC$. The tensors $\delta_{ab}^\ms{(n)}$ appearing in the boundary stress tensor have the form \eqref{ScalarPotential}, and therefore do not contribute to the conserved charges. 

Similarly, a tensor $\chi_{ab}$ admits a symmetric, transverse tensor potential if
\begin{equation}\label{TensorPotential}
\chi_{ab} = \DD^{2} \gamma_{ab} + 2\,\RR^\ms{(0)}_{acbd}\,\gamma^{cd} +
  (B - A)\,\left(\delta_{a}{}^{\alpha}\,\gamma_{\alpha b}+\delta_{b}{}^{\alpha}\,\gamma_{\alpha a}\right)
\end{equation}
for some symmetric $\gamma_{ab}$ with $\DD^{a}\gamma_{ab} = 0$. As in the case of the scalar potential, the divergence of the tensor $\chi_{ab}$ does not vanish. However, after commuting covariant derivatives and applying the contracted Bianchi identities, the divergence can be written in the form
\begin{gather}
 \DD^{b} \chi_{ab} = 2\,(B-A)\,\DD^{\alpha} \gamma_{\alpha a} ~.
\end{gather}
The current formed from $\chi_{ab}$ and the Killing vector $\xi_\ms{(0)}^{b}$ takes a form similar to \eqref{thetacurrent}
\begin{gather}
\chi_{ab} \, \xi_\ms{(0)}^{b} = 2\,\DD^{c}\left( \xi_\ms{(0)}^{b}\,\DD_{[c}\,\gamma_{a]b} + \gamma_{b[c} \DD_{a]}\,\xi_\ms{(0)}^{b} \right) + 2\,(B-A)\,\delta_{a}{}^{\alpha}\,\xi_\ms{(0)}^{b}\,\gamma_{\alpha b} ~.
\end{gather}
This current also has zero flux across $\CC$, and so it does not contribute to a conserved charge. Notice that an arbitrary tensor along $\Sigma_\ts{B}$ can be added to the right-hand side of \eqref{TensorPotential} (or \eqref{ScalarPotential}, for that matter) without changing this conclusion.

The equations of motion in appendix \ref{sec:EOMandBianchi} can be used to show that certain terms in the boundary stress tensor admit scalar and tensor potentials. Consider the tensor $\gamma_{ab}^\ms{(1)}$, defined in \eqref{gammadef}. The divergence of $\gamma_{ab}^\ms{(1)}$ vanishes by \eqref{AsympMom1}. Rewriting the equations of motion in terms of $\gamma_{ab}^\ms{(1)}$ gives
\begin{align}\nonumber
(d-2)\,\gamma_{ab}^\ms{(1)} + \frac{8\,(p-1)}{\alpha^{2}}\,\left(\delta_{a}{}^{\alpha} \, \gamma_{\alpha b}^\ms{(1)} + \delta_{b}{}^{\alpha} \, \gamma_{a \alpha}^\ms{(1)}\right)
 = & \dub \chi_{ab}[\gamma_{ab}^\ms{(1)}] - \theta_{ab}[h^\ms{(1)}] + \frac{16\,(p-1)}{\alpha\,(d-1)\,(d-2)}\,\theta_{ab}[\phi^\ms{(1)}] \\  \label{gamma1}
    & \dub + \frac{8\,(p-1)}{\alpha^{2}}\,\left(\delta_{a}{}^{\alpha} \, W_{\alpha b}^\ms{(1)}
 + \delta_{b}{}^{\alpha} \, W_{\alpha a}^\ms{(1)}\right) \\ \nonumber
  & \dub + \frac{8\,(p-1)}{\alpha^{2}}\,\delta_{a}{}^{\alpha}\,\delta_{b}{}^{\beta}\,
  \left(h_{\alpha\beta}^\ms{(1)} - h_{\alpha\beta}^\ms{(0)}\,h_{s}^\ms{(1)} + 2\,\alpha\,h_{\alpha\beta}^\ms{(0)}\,\phi^\ms{(1)} \right) ~.
\end{align}
Most of the terms on the right-hand side of this expression do not contribute to conserved charges, either because they admit potentials or because they have both indices along $\Sigma_\ts{B}$. The only exception occurs for asymptotic Killing fields $\xi^{\alpha}$ with a component along $\Sigma_\ts{B}$. In that case a term proportional to $u_\ms{(0)}^{a}\,W_{\alpha a}^\ms{(1)}\,\xi_\ms{(0)}^{\alpha}$ appears in the expression for the flux of the current $\gamma_{a\alpha}^\ms{(1)}\,\xi_\ms{(0)}^{\alpha}$ across ${\cal C}$\,\footnote{This term indicates the presence of an electric field along $\Sigma_\ts{B}$ at the first sub-leading order in the asymptotic expansion.}. We require that such a term is not present: $u_\ms{(0)}^{a}\,W_{\alpha a}^\ms{(1)} = 0$ for any timelike $u^{a}$. This condition ensures that terms proportional to $\gamma_{ab}^\ms{(1)}$ do not contribute to the conserved charges. Furthermore, this result and the analogous result for $\delta_{ab}^\ms{(1)}$ ensure that the integral of the first term in \eqref{QExpression} vanishes, which proves that the conserved charges are well-defined. 

Now consider the tensor $\gamma_{ab}^\ms{(2)}$
\begin{gather}
 \gamma_{ab}^\ms{(2)} = h_{ab}^\ms{(2)} - h_{ab}^\ms{(0)}\,h^\ms{(2)} + \frac{8\,(p-1)}{\alpha\,(d-1)^{2}}\,h_{ab}^\ms{(0)}\,\phi^\ms{(2)} ~.
\end{gather}
In this case the equations of motion only constrain the components of $\gamma_{ab}^\ms{(2)}$ with at least one index along $\Sigma_\ts{B}$
\begin{align}\label{gamma2}
\frac{8\,(p-1)}{\alpha^{2}}\,\left(\delta_{a}{}^{\alpha} \, \gamma_{\alpha b}^\ms{(2)}
 + \delta_{b}{}^{\alpha} \, \gamma_{\alpha a}^\ms{(2)}\right) =
 & \dub \chi_{ab}[\gamma_{ab}^\ms{(2)}] - \theta_{ab}[h^\ms{(2)}] + \frac{16\,(p-1)}{\alpha\,(d-1)^{2}}\,\theta_{ab}[\phi^\ms{(2)}] \\ \nonumber
    & \dub + \frac{8\,(p-1)}{\alpha^{2}}\,\left(\delta_{a}{}^{\alpha} \, W_{\alpha b}^\ms{(2)}
 + \delta_{b}{}^{\alpha} \, W_{\alpha a}^\ms{(2)}\right) \\ \nonumber
  & \dub + \frac{8\,(p-1)}{\alpha^{2}}\,\delta_{a}{}^{\alpha}\,\delta_{b}{}^{\beta}\,
  \left(h_{\alpha\beta}^\ms{(2)} - h_{\alpha\beta}^\ms{(0)}\,h_{s}^\ms{(2)} + 2\,\alpha\,h_{\alpha\beta}^\ms{(0)}\,\phi^\ms{(2)} \right) ~.
\end{align}
Thus, the term in $T_{ab}^\ms{(2)}$ proportional to $\gamma_{ab}^\ms{(2)}$ can contribute to the conserved charges. In fact, we have shown that this is the \emph{only} term that contributes to the conserved charges, which can now be expressed as
\begin{equation}\label{FinalQExpression}
  Q[\xi] = (d-1)\,\int_{\CC} \nts d^{\,d-1}x \,\sqrt{h_\ms{\CC}^\ms{(0)}}\,u^{a}_\ms{(0)}\,\gamma_{ab}^{(2)}\,\xi^{b}_\ms{(0)} ~.
\end{equation}

\vskip1em
\noindent{\bf Conserved Charges are Independent of Free Parameters in the Action}
\vskip.5em

\noindent In the special case $\alpha = \pm 2\,(p-1)/(d-1)$ the boundary stress tensor contains terms proportional to the free parameter $c_1$. However, all of these terms are of the sort described in the previous section -- they admit scalar or tensor potentials and therefore do not contribute to the conserved charges. So, for theories that admit a one-parameter family of two-derivative counterterms, the conserved charges are independent of the choice of action. This is also true for actions that include higher-derivative counterterms, as we will now show. 

Consider the counterterm action discussed in section \ref{sec:HDC}, which includes a four derivative term proportional to $\RR^{2}$. This leads to a one-parameter family of actions for generic values of $\alpha$, and a corresponding family of boundary stress tensors. We will now show that these stress tensors differ only by terms that admit potentials, so that the conserved charges are independent of the choice of action. The difference between $P_{ab}$ obtained from \eqref{HDCT} and $P_{ab}$ obtained from \eqref{CT} is
\begin{align}
\Delta P_{ab} = & \,\, \frac{1}{2}\,b_3\,\beta^{4}\,\RR_{(0)}^{\,2}\,h_{ab}\,e^{-\frac{\alpha}{p-1}\,\phi} + 2\,b_3\,\beta^{2}\,\RR^{(0)}\,e^{\frac{\alpha}{p-1}\,\phi}\,\left(\RR_{ab} - \frac{1}{2}\,h_{ab}\,\RR \right) \\ \nonumber
& \,\, -2\,b_3\,\beta^{2}\,\RR^{(0)}\,\left(D_{a}D_{b}\left(e^{\frac{\alpha}{p-1}\,\phi}\right)  - h_{ab}\,D^{2}\left(e^{\frac{\alpha}{p-1}\,\phi}\right)\right) \\ \nonumber
& \,\, + b_3 \, e^{\frac{3\,\alpha}{p-1}\,\phi} \left(\frac{1}{2}\,h_{ab}\,\RR^{2} - 2\,\RR\,\RR_{ab} \right) + 2\,b_3\,D_{a}D_{b}\left(e^{\frac{3\,\alpha}{p-1}\,\phi}\,\RR \right) \\ \nonumber
& \,\, - 2\,b_3\,h_{ab}\,D^{2}\left(e^{\frac{3\,\alpha}{p-1}\,\phi}\,\RR \right) ~,
\end{align}
with $b_3$ an arbitrary real number. An extended calculation involving the asymptotic expansions of all the terms gives
\begin{align}
\Delta P_{ab} = & \,\, b_3 \, \beta^{\,3}\,\rho^{3-d}\,\left(\frac{4\,\alpha}{p-1}\,\RR^{(0)}\,\theta_{ab}[\phi^{(1)}] +\theta_{ab}[\RR^{(1)}-\RR_{cd}^{(0)}\,h^{cd}_{(1)}]\right) \\ \nonumber
& \, \, + b_3 \, \beta^{\,3}\,\rho^{2-d}\,\left(\frac{4\,\alpha}{p-1}\,\RR^{(0)}\,\theta_{ab}[\phi^{(2)}] +\theta_{ab}[\RR^{(2)}-\RR_{cd}^{(0)}\,h^{cd}_{(2)}]\right) + \OO(\rho^{1-d}) ~.
\end{align}
Thus, when the higher derivative counterterm proportional to $\RR^{2}$ is included, the only affect on the boundary stress tensor is to generate terms admitting scalar potentials. Such terms do not change the conserved charges of the theory. Similar results were also obtained for higher derivative counterterms proportional to $\RR^{ab}\,\RR_{ab}$, $\RR\,\FF^{2}$, $\RR^{ab}\,\FF_{a}{}^{c}\,\FF_{bc}$, and $\RR^{k}$ for $k> 2$. We conclude that this result holds in general: the conserved charges are always independent of free parameters that appear in the action. This is true for any value of $\alpha$, and for actions containing boundary terms with any number of derivatives.

\section{Examples}
\label{sec:Examples}

A nice application of the techniques developed in the preceding sections is the holographic renormalization of asymptotically linear dilaton spacetimes that arise in type II supergravity theories. The field content in the NS-NS sector is a metric $g_{ab}$, a dilaton $\phi$, and a two-form potential ${\bf B}$ with field strength ${\bf H} = \textrm{d} {\bf B}$. In addition, there is another two-form potential ${\bf C}$ in the R-R sector, with field strength ${\bf F} = \textrm{d} {\bf C}$. The bulk part of the two-derivative action for these fields is
\begin{equation}
 L = \frac{1}{2\,\kappa^{2}}\,\sqrt{g}\,\left[R - \frac{1}{2}\,\left(\nabla \phi \right)^{2} - \frac{1}{12}\,e^{-\phi}\,H^{abc}\,H_{abc} -\frac{1}{12}\,g_{s}^{\,2}\,e^{\,\phi}\,F^{abc}\,F_{abc}\right]
\end{equation}
with $2\,\kappa^{2}$ = $2\,\kappa_{0}^{\,2}\,g_{s}^{\,2}$ and $2\,\kappa_{0}^{\,2} = (2\pi)^{7}\,(\alpha')^{4}$. This is the same form as \eqref{bulkaction}, with $\alpha = -1/2$ for the NS-NS coupling and $\alpha = 1/2$ for the R-R coupling. The R-R field strength appears in the Einstein frame action with an additional factor of $g_{s}^{\,2}$, since it does not couple to the dilaton in string frame.

\subsection{Thermally Excited NS5-Branes}
\label{sec:NS5}

A counterterm action based on the $\hK$ construction was used in \cite{Marolf:2007ys} to holographically renormalize the NS-NS sector of the action, for spacetimes asymptotic to a stack of non-extremal NS 5-branes. As a first application of our results we repeat that analysis using the explicit boundary counterterms of section \ref{sec:OC}. Notice that $\alpha=-1/2$ corresponds to the special case $\alpha = - 2 (p-1)/(d-1)$, since $p=3$ and $d=9$. This means that there is a one parameter family of holographically renormalized, two-derivative actions to consider. In the last section we showed that all such actions yield the same conserved charges. For the present example, it turns out that the stress tensor itself is also independent of the choice of action. Thus, the results given below hold for any value of the free parameter $c_1$.

Consider a stack of $N$ coincident non-extremal NS 5-branes. The field strength is
\begin{equation}\label{NS5flux}
  H_{\alpha\beta\gamma} = 2\,N\,\alpha'\,\eps^\ms{\,\Omega}_{\alpha\beta\gamma}~,
\end{equation}
with $\eps^\ms{\Omega}_{\alpha\beta\gamma}$ the volume form on a unit three sphere in the directions transverse to the NS 5-branes. The near-horizon solution describing thermal fluctuations of the fields is \cite{Narayan:2001dr}
\begin{gather}\label{NearHorizonNS5}
ds^{2} = \left(\frac{N\,\alpha'}{r^{2}}\right)^{-\tfrac{1}{4}}\left( -\left( 1 - \frac{r_{h}^{\,2}}{r^{2}}\right)\,dt^{2} + d \vec{x}_{5}^{\,2} \right) + \left(\frac{N\,\alpha'}{r^{2}}\right)^{\tfrac{3}{4}} \left( \left(1-\frac{r_{h}^{\,2}}{r^{2}}\right)^{-1} \nts dr^{2} + r^{2}\,d\Omega_{3}^{\,2} \right) \\
\phi  = \frac{1}{2}\,\log\left(\frac{N\,\alpha'}{r^{2}}\right)
\end{gather}
where $r_{h}$ is the non-extremality parameter. To express this solution in the form \eqref{metricBC} and \eqref{dilatonBC}, we make two changes of coordinates. First, define $r = r_h \, \cosh \sigma$ and $\ell^{\,2} = N\,\alpha'$, so that the solution becomes
\begin{gather}\label{coord1}
ds^{2}  = \left(\frac{r_h}{\ell}\,\cosh \sigma \right)^{1/2} \bigg[\, \ell^{\,2}\,\left( d\sigma^{2} + d\Omega_{3}^{\,2} \right)  - \tanh^{2} \sigma \, d t^{2} + d \vec{x}_{5}^{\,2} \bigg] \\
\phi = - \log \left( \frac{r_h}{\ell}\,\cosh \sigma  \right) ~.
\end{gather}
Now convert to asymptotic coordinates via
\begin{equation}\label{coord2}
 \sigma = \log \left( \frac{\rho^{\,4}}{2^{7}\,r_h\,\ell^{\,3}} \right) + \frac{4^{7}\,r_{h}^{\,2}\,\ell^{\,6}}{7\,\rho^{\,8}} + \OO(\rho^{\,-9})~,
\end{equation}
and rescale the worldvolume coordinates as
\begin{equation}
t = 4\,\ell\,\tau \quad\quad\quad \vec{x}_{5} = 4\,\ell\,\vec{y}_{5} ~.
\end{equation}
In these coordinates the solution becomes
\begin{gather}\label{metricNS5branes}
ds^{2} = d \rho^{\,2} + \rho^{\,2}\,\bigg[ -\left(1 - \frac{6}{7}\,\frac{\mu}{\rho^{8}} + \ldots\right)\,d\tau^{2} + \left(1+\frac{1}{7}\,\frac{\mu}{\rho^{8}} + \ldots \right)\,\left(d\vec{y}_{5}{}^{2} + \frac{1}{16}\,d\Omega_{3}{}^{2}\right)\bigg] \\ \label{dilatonNS5branes}
\phi = - 4\,\log \left(\frac{\rho}{4\,\ell} \right) - \frac{2}{7}\,\frac{\mu}{\rho^{8}} + \ldots \\ \label{FSNS5branes}
H_{\alpha\beta\gamma} = 2\,\ell^{\,2}\,\eps^\ms{\,\Omega}_{\alpha\beta\gamma}
\end{gather}
where $\mu$ is related to the non-extremality parameter $r_h$ and the length scale $\ell$ by
\begin{gather}\label{MassParam}
\mu = 4^{8}\,r_{h}^{\,2}\,\ell^{\,6}
\end{gather}
At this point, the components of the stress tensor can be directly obtained from the asymptotic expansions of the fields.

The first thing to notice is that $h_{ab}^\ts{(1)} = 0$ and $\phi^\ts{(1)}=0$, and there are no sub-leading terms in the field strength. The boundary stress tensor can be read off from the solution \eqref{metricNS5branes}. The only non-vanishing component is
\begin{equation}
T_{\tau\tau}^\ts{(2)} = \frac{4\,\mu}{\kappa^{\,2}} ~.
\end{equation}
The conserved charge associated with the asymptotic Killing vector $\partial_{\tau}$ is then given by
\begin{align}
Q[\partial_{\tau}] = \int_{\CC}\nts d^{\,8}x\,\sqrt{h_{\CC}^{_{(0)}}}\,u_\ms{(0)}^{\tau}\,T_{\tau\tau}^{(2)}\,u_\ms{(0)}^{\tau} ~.
\end{align}
Taking $\CC$ to be a surface of constant $\tau$, with $\omega_{\alpha\beta}$
the metric on the unit 3-sphere, we have from \eqref{metricNS5branes}
\begin{equation}
\sqrt{h_{\CC}^{_{(0)}}} = \frac{1}{4^{3}}\,\sqrt{\omega} \quad \quad u^{t} = 1 ~.
\end{equation}
This gives
\begin{equation}\label{PreMassNS}
Q[\partial_{\tau}] = \frac{\mu}{4^{2}\kappa^{2}}\,{\textrm {A}}(\Omega_{3})\,{\textrm {Vol}_{y}}(\BR^{5})~,
\end{equation}
where $\textrm{A}(\Omega_{3})$ is the area of the unit 3-sphere, and the volume factor represents the integral over the rescaled coordinates $\vec{y}_5$ on the worldvolume of the branes. Using the definition \eqref{MassParam} and transforming back to the original worldvolume coordinates $t = 4\,\ell\,\tau$ and $\vec{x}_5 = 4\,\ell\,\vec{y}_5$, we then have
\begin{gather}\label{Mass1}
Q[\partial_{t}] = \frac{r_{h}^{\,2}}{\kappa^{2}}\,{\textrm {A}}(\Omega_{3})\,{\textrm {Vol}_{x}}(\BR^{5})~,
\end{gather}
which we interpret as the mass (internal energy) $M$ of the spacetime. 

As a test of this result we can check it against the first law. Continuing the solution \eqref{NearHorizonNS5} to Euclidean time, regularity at the horizon implies that the Hawking temperature is
\begin{equation}
T_{H} = \frac{1}{2\pi\sqrt{N\,\alpha'}} ~.
\end{equation}
This result is independent of the non-extremality parameter, which leads to a vanishing free energy. The entropy of the system is one-quarter of the area of the horizon in Planck units
\begin{equation}
  S = \frac{A_{h}}{4\,G} = \frac{1}{\kappa^{2}}\,2\pi\sqrt{N\,\alpha'}\,r_{h}^{\,2}\,{\textrm {A}}(\Omega_{3})\,{\textrm {Vol}_{x}}(\BR^{5})
\end{equation}
Comparing with the result \eqref{Mass1}, we have
\begin{equation}
 dM = T_{H}\,dS ~,
\end{equation}
in agreement with the first law of thermodynamics.

\subsection{D5-Branes}
\label{sec:D5}

The situation is similar for a stack of near-extremal D5-branes, except that the solution contains additional factors of the string coupling $g_s$. The R-R field strength is
\begin{equation}\label{D5flux}
  F_{\alpha\beta\gamma} = 2\,N\,\alpha'\,\eps_{\alpha\beta\gamma} ~,
\end{equation}
and the near-horizon solution is
\begin{gather}\label{NearHorizonD5}
ds^{2} = \left(\frac{g_s N \alpha'}{r^{2}}\right)^{-\frac{1}{4}}\left( -\left( 1 - \frac{r_{h}^{\,2}}{r^{2}}\right)\,dt^{2} + d \vec{x}_{5}^{\,2} \right) + \left(\frac{g_s N \alpha'}{r^{2}}\right)^{\tfrac{3}{4}} \left( \left(1-\frac{r_{h}^{\,2}}{r^{2}}\right)^{-1} \nts dr^{2} + r^{2}\,d\Omega_{3}^{\,2} \right) \\
\phi  = \frac{1}{2}\,\log\left(\frac{g_s N \alpha'}{r^{2}}\right) ~.
\end{gather}
Using the same coordinate transformations as in the previous section, the solution for a stack of D5-branes can be put in the form
\begin{gather}\label{metricD5branes}
ds^{2} = d \rho^{\,2} + \rho^{\,2}\,\bigg[ -\left(1 - \frac{6}{7}\,\frac{\bar{\mu}}{\rho^{8}} + \ldots\right)\,d\tau^{2} + \left(1+\frac{1}{7}\,\frac{\bar{\mu}}{\rho^{8}} + \ldots \right)\,\left(d\vec{y}_{5}{}^{2} + \frac{1}{16}\,d\Omega_{3}{}^{2}\right)\bigg] \\ \label{dilatonD5branes}
\phi = - 4\,\log \left(\frac{\rho}{4\,\bar{\ell}} \right) - \frac{2}{7}\,\frac{\bar{\mu}}{\rho^{8}} + \ldots \\ \label{FSD5branes}
F_{\alpha\beta\gamma} = 2\,\frac{\,\bar{\ell}^{\,2}}{g_{s}}\,\eps^\ms{\,\Omega}_{\alpha\beta\gamma} ~,
\end{gather}
where $\bar{\ell}$ and $\bar{\mu}$ are now defined as
\begin{gather}
\bar{\ell} = \sqrt{g_s \, N\, \alpha'} \quad \quad \quad \bar{\mu} = 4^{8}\,r_{h}^{\,2}\,\bar{\ell}^{\,6} ~.
\end{gather}
The $\tau$-$\tau$ component of the boundary stress tensor is now
\begin{equation}
T_{\tau\tau}^{(2)} = \frac{4\,\bar{\mu}}{\kappa^{\,2}} ~,
\end{equation}
so the conserved charge associated with the Killing vector $\partial_{\tau}$ is
\begin{equation}\label{PreMassD}
Q[\partial_{\tau}] = \frac{\bar{\mu}}{4^{2}\kappa^{2}}\,{\textrm {A}}(\Omega_{3})\,{\textrm {Vol}_{y}}(\BR^{5})~.
\end{equation}
This differs from the result \eqref{PreMassNS} for the NS 5-branes by a factor of $g_{s}^{\,3}$. However, in terms of the coordinates $t = 4\,\bar{\ell}\,\tau$ and $\vec{x}_{5} = 4\,\bar{\ell}\,\vec{x}_{5}$ appearing in \eqref{NearHorizonD5}, we obtain the same result as before\,\footnote{The results \eqref{Mass1} and \eqref{Mass2} are the same, but the coordinates used in these two examples are not.}
\begin{equation}\label{Mass2}
M = \frac{r_{h}^{\,2}}{\kappa^{\,2}}\,{\textrm{A}(\Omega_3)}\,{\textrm{Vol}_{x}}(\BR^{5})~.
\end{equation}
It is straightforward to show, using the same arguments as the previous section, that the first law of thermodynamics is upheld. Both the Hawking temperature and the entropy depend on $g_s$ in this case, but that this dependence cancels out when they are multiplied.

\section{Discussion}
\label{sec:Discussion}

The main lesson of this paper is that a well-defined variational problem is the key physical principle guiding the introduction of surface terms in the action. In the case of asymptotically flat or asymptotically AdS spacetimes the potential boundary counterterms are quite restrictive. But for ALD gravity the boundary terms are not necessarily unique -- it is possible to construct families of actions, all of which have a well-defined variational principle. However, these actions all have the same value on-shell, and they all yield the same conserved charges.

For ALD gravity one can introduce a set of counterterms based on the $\hK$ approach. This makes contact with similar constructions in asymptotically flat gravity. Specifically, the $\alpha \to \infty$ limit in \eqref{DefiningEquation} recovers the definition of $\hK_{ab}$ used previously \cite{Mann:2005yr, Mann:2006bd}. However the definition of $\hK$ for ALD gravity is inherently ambiguous because of relations between the boundary data that are not present in AF gravity. Regardless of which definition is used, the $\hK$ construction does not lead to an action with a well-defined variational principle. Additional surface terms must be included in the action, as was first pointed out in \cite{Marolf:2007ys}. This suggests an alternate approach in which $\hK$ -- whose implicit definition makes certain calculations cumbersome -- is eliminated from the boundary counterterm action. Assuming a set of surface terms with no more than two derivatives, in section \ref{sec:OC} we arrived at an action of the form
\begin{align}\label{FinalAction}
I = & \dub \frac{1}{2\,\kappa^{\,2}}\,\int_{\MM} \nts \nts d^{\,d+1}x \sqrt{g} \, \left( R - 
  \frac{4}{d-1} \, \nabla^{\mu} \phi \nabla_{\mu} \phi - \frac{1}{2 \cdot p!} \, e^{2 \alpha 
  \phi} \, F^{\,a_1 \ldots a_p} F_{a_1 \ldots a_p}\right) \\ \nonumber
    & \dub \dub +  \frac{1}{\kappa^{\,2}}\,\int_{\dM} \bns \nts d^{\,d} x \,\sqrt{h} \, 
  \bigg( K  +c_{0}\,e^{-\frac{\alpha}{p-1}\,\phi} + e^{\frac{\alpha}{p-1}\,\phi} \Big(
  c_{1}\,\RR + c_{2} \,e^{2\,\alpha\,\phi}\,\FF^{\,a_\ms{1}\ldots a_\ms{p}} \,\FF_{a_\ms{1}\ldots a_\ms{p}}\Big) \bigg) ~.
\end{align}
For most values of $\alpha$, requiring a well-defined variational principle determines the coefficients of the surface terms uniquely \eqref{uniqueci}. In the special case $\alpha = \pm 2(p-1)/(d-1)$ the conditions that determine the $c_i$ become linearly dependent, and there is a one-parameter family of actions \eqref{SpecialAlpha} with a well-defined variational principle. 

Allowing boundary counterterms with more than two derivatives reveals the existence of families of actions that contain an arbitrary number of free parameters. This is quite unlike the situation in AAdS gravity, where counterterms with more derivatives are less relevant in the asymptotic expansion. Higher derivative counterterms in ALD gravity may be dressed with factors of the dilaton so that they are just as relevant as their lower-derivative counterparts. However, we were able to show that both the on-shell action and the conserved charges are independent of the choice of counterterms. Quantities like the boundary stress tensor do, in principle, depend on the choice of action, though in the examples we considered in section \ref{sec:Examples} this was not the case. It would be interesting to find an example where this scheme dependence could be examined in the context of a dual holographic theory.

Of course, there are various reasons for restricting our attention to a finite number of counterterms. For example, a particular supergravity theory may be of interest as the low energy limit of string theory. At lowest order in $\alpha'$ one would expect terms with no more than two derivatives in either the bulk or boundary part of the action. Even in that case there are still examples that admit a one-parameter family of counterterms. It is worth pointing out a similar situation in two dimensions: the low energy limit of type 0 supergravity also seems to admit a one parameter family of boundary counterterms, but requiring the full action to respect Buscher duality \cite{Buscher:1987qj} identifies a unique action \cite{Davis:2004xi}. Perhaps T-duality or another symmetry can be used to identify a unique action for the ten dimensional type II supergravity examples we considered in section \ref{sec:Examples}.

The constructions presented in this paper suggest a number of generalizations. The two most obvious are additional field content and a broader class of boundary conditions. For instance, one might consider additional scalar fields or field strengths with support at spatial infinity. At the same time, the boundary conditions could be weakened by allowing terms that fall off more slowly as $\rho \to \infty$. In particular, it would be interesting to investigate boundary conditions associated with more complicated (and less symmetric) brane configurations than the simple stacks of branes analyzed in section \ref{sec:Examples}.

Finally, there is a puzzling aspect to our results that we have not found a satisfactory explanation for. The fall-off conditions in section \ref{sec:Prelim} allow for terms like $h_{ab}^\ms{(1)}$ and $\phi^\ms{(1)}$ at relative order $\OO(\rho^{2-d})$ in the asymptotic expansion. As shown in section \ref{sec:LRF}, these terms do not contribute to the conserved charges. Yet they seem to have non-trivial equations of motion, and so we would expect that they are dynamical. This is certainly the case in AF gravity, where the analogous terms encode useful information about the spacetime and contribute to the conserved charges associated with asymptotic translations. It is possible that we have simply missed some consequence of the equations of motion or Bianchi identities that force these terms to vanish for ALD gravity, as is the case for the two examples in section \ref{sec:Examples}. On the other hand, it may be that these terms are relevant for some aspect of the theory that is simply not apparent when linearizing the equations of motion.

\vskip1em
\noindent{\bf Acknowledgments}
\vskip.5em
The authors would like to thank Don Marolf, Amitabh Virmani, and Oleg Lunin for useful correspondence and conversations. Research at Perimeter Institute is supported by the Government of Canada
through Industry Canada and by the Province of Ontario through the Ministry of Research
\& Innovation. R. Mann would also like to thank the Kavli Institute for Theoretical Physics, where part of this work was carried out, and the Fulbright Foundation and the Natural Sciences and Engineering Research Council of Canada for financial support.

\appendix

\section{Hypersurfaces}
\label{sec:Hyper}

This appendix collects a number of useful results concerning the embedding of a
timelike, $d$-dimensional hypersurface $\Sigma \subset \MM$. The surface is described
locally by a spacelike, outward-pointing vector $n^{a}$ with unit norm. The induced metric on $\Sigma$ is given by the pullback of the bulk metric: $h_{ab} = g_{ab}|_\ms{\Sigma}$. This can be expressed in terms of $g_{ab}$ and the normal vector as
\begin{equation}
  h_{ab} = g_{ab} - n_{a} \, n_{b} ~.
\end{equation}
A tensor on $\MM$ can be projected onto $\Sigma$ by appropriate contraction of its indices with
$h_{a}{}^{b}$. The complete projection of a tensor is denoted by
\begin{equation}\label{projection}
 \perp T^{a \ldots}{}_{b \ldots}  = h^{a}{}_{c} \ldots h_{b}{}^{d}
 \ldots T^{c\ldots}{}_{d\ldots}
\end{equation}
where indices are lowered and raised using $g_{ab}$ and its inverse. If a tensor $V$ on $\MM$ is invariant under projection of all its indices then it is said to be parallel to $\Sigma$. The projection of the bulk covariant derivative acting on a tensor $V$ that is parallel to $\Sigma$ defines a covariant derivative $D_{a}$ on $\Sigma$
\begin{equation}
  D_{c} V^{a\ldots}{}_{b\ldots} \defeq \dub \perp \nabla_{c} V^{a\ldots}{}_{b\ldots} ~.
\end{equation}
The covariant derivative $D$ is compatible with the induced metric on $\Sigma$. The commutator of two such covariant derivatives acting on a vector parallel to $\Sigma$ defines the
intrinsic curvature of $(\Sigma,h)$
\begin{equation}
  \left[D_{a}, D_{b} \right] \, V^{c} = \RR^{c}{}_{d a b} \,
    V^{d} ~.
\end{equation}
The intrinsic Ricci tensor and scalar curvature are given by the appropriate contractions of the Riemann tensor
\begin{equation}
   \RR_{ab} = \RR^{c}{}_{a c b} \quad \quad \RR = \RR^{a}{}_{a} ~.
\end{equation}
In addition to the intrinsic curvatures, we also need the extrinsic curvature associated
with the embedding of $\Sigma$. This given by the change in the normal
vector projected onto $\Sigma$
\begin{equation}\label{ExtrinsicCurvatureDefn}
  K_{ab} = \,\,\perp \nabla_{a} \,n_{b} =
    \,\frac{1}{2} \, \LL_{n} h_{ab} ~,
\end{equation}
where $\LL_{n}$ is the Lie derivative along $n^{a}$.

There are a number of identities -- used extensively throughout the paper -- that relate bulk curvature tensors at $\Sigma$ to the intrinsic and extrinsic curvatures described above. The projections of the bulk Riemann tensor along and normal to $\Sigma$ are
\begin{align}\label{GC}
    \bot R_{c a d b} \, = & \dub \RR_{c a d b} - K_{c d}
    K_{a b} + K_{a d} K_{b c} \\ \raisebox{14pt}{} \label{GC2}
    \bot \big( R_{c a d b} \,n^{c} \big) \, = & \dub D_{b} K_{ad} -
    D_{d} K_{ab} \\ \raisebox{14pt}{} \label{GC3}
    \bot \big(R_{c a d b} \,n^{c}  n^{d}\big) \, = & \, - \LL_{n} \,
    K_{ab} + K_{a}{}^{c} \, K_{cb} + D_{a} a_{b} - a_{a} a_{b} ~.
\end{align}
It follows that the projections of the bulk Ricci tensor are
\begin{align}\label{Ricci}
 \bot \left( R_{ab} \right) \, = & \,\, \RR_{ab} + D_{a} a_{b} - a_{a} a_{b} -
 \LL_{n} \, K_{ab} - K \, K_{ab} + 2 K_{a}{}^{c} \, K_{bc} \\ \raisebox{14pt}{}
 \bot \left( R_{ab} \,n^{a}  \right) \, =  & \,\, D^{a} K_{ab} - D_{b} K
 \\ \raisebox{14pt}{}
 R_{ab} \,n^{a} n^{b} \, = & \raisebox{10pt}{\,} - \LL_{n} \, K - K^{ab} \, K_{ab} + D_{b} a^{b}
 -a_{b} \,a^{b}
\end{align}
Finally, the bulk Ricci scalar evaluated at $\Sigma$ can be expressed in terms of intrinsic and extrinsic quantities as
\begin{align}
 R \, = & \raisebox{10pt}{\,\,} \RR - K^2 - K^{ab}\,K_{ab} - 2\,\LL_{n} \, K + 2 \, D_{b} a^{b} - 2 \,
 a_{b} \,a^{b} ~.
\end{align}
The vector $a^{b}$ that appears in these expressions is defined as $a^{b} = n^{a} \nabla_{a} n^{b}$. It is straightforward to show that this vector vanishes for a constant $\rho$ surface embedded in a spacetime with metric \eqref{metricBC}.

\section{Asymptotic Expansions}
\label{sec:AsympExp}

Most of the calculations in this paper rely on asymptotic expansions of functions of the fields, such as curvature tensors or the equations of motion. The purpose of this appendix is to establish notation and collect a few essential results.

Terms in the asymptotic expansion are labeled by a numerical subscript or superscript -- the position of the labels is not important. The leading terms carry a `(0)', while next-to-leading and next-to-next-to-leading order terms carry a `(1)' and `(2)', respectively. For example, the expansion of the $d$-dimensional metric ($d>3$) was given in \eqref{metricBC} as
\begin{gather}\label{BndyMetricExpansion}
h_{ab} = \rho^{\,2}\,h_{ab}^\ms{(0)} + \rho^{\,4-d} \, h_{ab}^\ms{(1)} + \rho^{\,3-d}\,h_{ab}^\ms{(2)} + \ldots
\end{gather}
and the first few terms in the expansion of the inverse metric are
\begin{equation}\label{invmetexp}
  h^{ab} = \rho^{-2} \, h^{ab}_\ms{(0)} - \rho^{-d}\,h^{ab}_\ms{(1)} - \rho^{-(1+d)} \,h^{ab}_\ms{(2)} + \ldots ~.
\end{equation}
Indices on tensors that carry a numerical label are lowered and raised using $h_{ab}^\ms{(0)}$ and its inverse $h^{ab}_\ms{(0)}$. The trace of $h_{ab}^\ms{(n)}$ with respect to the metric $h^{ab}_\ms{(0)}$ is denoted by $h^\ms{(n)}$. Such terms appear, for instance, in the asymptotic expansion for the square-root of the determinant of the metric
\begin{equation}
   \sqrt{h} = \sqrt{h^\ms{(0)}}\,\left(\rho^{\,d} + \frac{1}{2}\,\rho^{\,2}\,h^\ms{(1)} + \frac{1}{2}\,\rho\,h^\ms{(2)} + \ldots \right)
\end{equation}
Only the first three orders of the asymptotic expansions will be relevant in this paper. Higher-order terms are either denoted by `$\ldots$', or simply dropped.

Using the notation described above, the intrinsic Ricci tensor on $\Sigma$ takes the form
\begin{equation}\label{RicciTensorExpansion}
  \RR_{ab} = \RR_{ab}^\ms{(0)} + \rho^{\,2-d}\,\RR_{ab}^\ms{(1)} + \rho^{\,1-d}\,\RR_{ab}^\ms{(2)} + \ldots
\end{equation}
with $\RR_{ab}^\ms{(0)}$ the Ricci tensor for $h_{ab}^\ms{(0)}$. The terms $\RR_{ab}^\ms{(1)}$ and $\RR_{ab}^\ms{(2)}$ may be obtained from the usual expressions
\begin{align}
  \RR_{ab}^\ms{(1)} = & \,\,\frac{1}{2} \, \left( \DD^{c} \DD_{a} h_{bc}^\ms{(1)}
    + \DD^{c} \DD_{b} h_{ac}^\ms{(1)} - \DD_{a}\DD_{b} h^\ms{(1)} - \DD^{2} h_{ab}^\ms{(1)}\right) \\
  \RR_{ab}^\ms{(2)} = & \,\,\frac{1}{2} \, \left( \DD^{c} \DD_{a} h_{bc}^\ms{(2)}
    + \DD^{c} \DD_{b} h_{ac}^\ms{(2)} - \DD_{a}\DD_{b} h^\ms{(2)} - \DD^{2} h_{ab}^\ms{(2)}\right)
\end{align}
where $\DD_{a}$ is covariant derivative compatible with the metric $h_{ab}^\ms{(0)}$. Combining \eqref{RicciTensorExpansion} with the expression \eqref{invmetexp} gives the asymptotic expansion of the intrinsic Ricci scalar on $\Sigma$
\begin{align}
  \RR = \rho^{-2}\,\RR^\ms{(0)} + \rho^{-d}\,\left(h^{ab}_\ms{(0)}\,\RR_{ab}^\ms{(1)} - h^{ab}_\ms{(1)}\,\RR^{ab}_\ms{(0)} \right) + \rho^{-1-d}\,\left(h^{ab}_\ms{(0)}\,\RR_{ab}^\ms{(2)} - h^{ab}_\ms{(2)}\,\RR^{ab}_\ms{(0)} \right) + \ldots ~.
\end{align}
The asymptotic expansion of the extrinsic curvature is given by the Lie derivative of \eqref{BndyMetricExpansion} along $n^{a} = \delta^{a}{}_{\rho}$
\begin{gather}
K_{ab} = \frac{1}{2}\,n^{c}\,\partial_{c} h_{ab} = \rho\,h_{ab}^\ms{(0)} - \frac{d-4}{2}\,\rho^{\,3-d}\,h_{ab}^\ms{(1)} - \frac{d-3}{2}\,\rho^{\,2-d}\,h_{ab}^\ms{(2)} + \ldots ~.
\end{gather}
The expansion for the trace of the extrinsic curvature follows from this result and \eqref{invmetexp}
\begin{gather}
K = d\,\rho^{-1} - \frac{d-2}{2}\,\rho^{\,1-d}\,h^\ms{(1)} - \frac{d-1}{2}\,\rho^{-d}\,h^\ms{(2)} + \ldots ~.
\end{gather}
These basic results can be combined to obtain the various expansions used throughout the paper.

\section{Equations of Motion and Bianchi Identities}
\label{sec:EOMandBianchi}

The equations of motion \eqref{EinsteinEOM}, \eqref{DilatonEOM}, and \eqref{MaxwellEOM} can be solved order-by-order in the asymptotic expansion. This gives a set of equations satisfied by the coefficients in the expansions \eqref{metricBC}, \eqref{dilatonBC}, and \eqref{fluxBC}. The derivation of these equations involves a great deal of algebra and is not particularly illuminating; they are presented here so that they may be referred to from the main text.

We begin with the components of the Einstein equation \eqref{EinsteinEOM}. The analog of the Hamiltonian constraint is obtained by contracting both indices with the normal vector $n^{a}$. This yields the following equations
\begin{align}\label{AsympHam1}
(d-2)(d-3)\, h^\ms{(1)} + \frac{8\,(p-1)^2}{\alpha^{2}\,(d-1)}\,h_{s}^\ms{(1)} - \frac{16\,(p-1)\,(d+p-3)}{\alpha\,(d-1)}\,\phi^\ms{(1)} - \frac{8\,(p-1)^2}{\alpha^{2}\,(d-1)}\,Y^\ms{(1)} = & \dub 0 \\ \label{AsympHam2}
(d-1)(d-2)\, h^\ms{(2)} + \frac{8\,(p-1)^2}{\alpha^{2}\,(d-1)}\,h_{s}^\ms{(2)} - \frac{16\,(p-1)\,(d+p-2)}{\alpha\,(d-1)}\,\phi^\ms{(2)} - \frac{8\,(p-1)^2}{\alpha^{2}\,(d-1)}\,Y^\ms{(2)}  = & \dub 0 ~,
\end{align}
where $h^{_{(n)}} = h^{ab}_{^\ms{(0)}}\,h_{ab}^{_{(n)}}$, $h^{_{(n)}}_{s} = h^{\alpha\beta}_{^\ms{(0)}}\,h_{\alpha\beta}^{_{(n)}}$, and $Y^{{(n)}}$ is defined as
\begin{equation}\label{YDef}
Y^\ms{(n)} = \frac{2}{Q\,p!}\,\eps^{\alpha_1 \ldots \alpha_p}\,F_{\alpha_1 \ldots \alpha_p}^\ms{(n)} ~.
\end{equation}
Similarly, the analog of the momentum constraint follows from contracting one index with $n^{a}$ and projecting the remaining index along $\Sigma$ with $h^{a}{}_{b}$.
\begin{align}\label{AsympMom1}
\dub \DD^{a} h_{ab}^\ms{(1)} - \DD_{b} h^\ms{(1)} = & \dub -\frac{8\,(p-1)}{\alpha\,(d-1)\,(d-2)}\,\DD_{b}\phi^\ms{(1)} - \frac{8\,(p-1)}{\alpha^{\,2}\,(d-2)}\,\delta_{b}^{\dub \alpha} \, Z_{\alpha}^\ms{(1)} \\
\label{AsympMom2}
\dub \DD^{a} h_{ab}^\ms{(2)} - \DD_{b} h^\ms{(2)} = & \dub -\frac{8\,(p-1)}{\alpha\,(d-1)^{2}}\,\DD_{b}\phi^\ms{(2)} - \frac{8\,(p-1)}{\alpha^{\,2}\,(d-1)}\,\delta_{b}^{\dub \alpha} \, Z_{\alpha}^\ms{(2)} ~.
\end{align}
The variable $Z_{\alpha}^{_{(n)}}$ in these equations is defined as
\begin{gather}\label{ZDef}
   Z_{\alpha}^\ms{(n)} = \frac{1}{Q\,(p-1)!}\,\eps_{\alpha\,\alpha_{2}\ldots \alpha_{p}}^\ms{(0)}\,t^{\alpha_{2}\ldots \alpha_{p}}_{(n)} ~.
\end{gather}
The remaining components of the Einstein equation are parallel to $\Sigma$.
\begin{align}\label{AsympEinstein1}
\RR_{ab}^\ms{(1)} = & \dub \left(\frac{d}{2} - \frac{4\,(p-1)^2}{\alpha^{\,2}\,(d-1) } \right)\,h_{ab}^\ms{(1)} - \frac{(d-2)}{2 }\,h_{ab}^\ms{(0)}\,h^\ms{(1)} - \frac{4\,(p-1)^{2}}{\alpha^{\,2}\,(d-1) }\,h_{ab}^\ms{(0)}\,\left(2\,\alpha\,\phi^\ms{(1)} -h_{s}^\ms{(1)} + Y^\ms{(1)}\right) \\ \nonumber
 & \dub + \frac{4\,(p-1)}{\alpha^{\,2}}\,\delta_{a}{}^{\alpha} \delta_{b}{}^{\beta}\,
 \left( h_{\alpha\beta}^\ms{(1)}-h_{\alpha\beta}^\ms{(0)}\,h_{s}^\ms{(1)} + 2\,\alpha\,h_{\alpha\beta}^\ms{(0)}\,\phi^\ms{(1)}
 \right) + \frac{4\,(p-1)}{\alpha^{\,2} }\,\left(\delta_{a}{}^{\alpha}\,W_{\alpha b}^\ms{(1)}
 + \delta_{b}{}^{\alpha}\,W_{\alpha a}^\ms{(1)}\right) \\ \label{AsympEinstein2}
\RR_{ab}^\ms{(2)} = & \dub (d-1)\,\left( 1 - \frac{4\,(p-1)^{2}}{\alpha^{\,2}\,(d-1)^2}\right) \,h_{ab}^\ms{(2)} - \frac{(d-1)}{2}\,h_{ab}^\ms{(0)}\,h^\ms{(2)} - \frac{4\,(p-1)^{2}}{\alpha^{\,2}\,(d-1) }\,h_{ab}^\ms{(0)}\,\left(2\,\alpha\,\phi^\ms{(2)} -h_{s}^\ms{(2)} + Y^\ms{(2)}\right) \\ \nonumber
 & \dub + \frac{4\,(p-1)}{\alpha^{\,2}}\,\delta_{a}{}^{\alpha} \delta_{b}{}^{\beta}\,
 \left( h_{\alpha\beta}^\ms{(2)}-h_{\alpha\beta}^\ms{(0)}\,h_{s}^\ms{(2)} + 2\,\alpha\,h_{\alpha\beta}^\ms{(0)}\,\phi^\ms{(2)}
 \right) + \frac{4\,(p-1)}{\alpha^{\,2} }\,\left(\delta_{a}{}^{\alpha}\,W_{\alpha b}^\ms{(2)}
 + \delta_{b}{}^{\alpha}\,W_{\alpha a}^\ms{(2)}\right) ~,
\end{align}
with $W_{\alpha\mu}^\ms{(n)}$ given by
\begin{gather}\label{WDef}
W_{\alpha\mu}^\ms{(n)} = \frac{1}{Q\,(p-1)!}\,\eps_{\alpha \beta_{2}\ldots \beta_{p}}^\ms{(0)}\,F_{\mu}^{(n)\beta_{2}\ldots \beta_{p}} ~.
\end{gather}
Taking the trace of equations \eqref{AsympEinstein1} and \eqref{AsympEinstein2} with $h^{ab}_\ms{(0)}$ gives
\begin{align}\label{TraceAsympEinstein1}
h^{ab}_\ms{(0)} \, \RR_{ab}^\ms{(1)} = & \dub - \left(\frac{d\,(d-3)}{2 }+ \frac{4\,(p-1)^2}{\alpha^{\,2}\,(d-1) } \right)\,h^\ms{(1)} + \frac{4\,(p-1)^{2}}{\alpha^{\,2}\,(d-1) }\,h_{s}^\ms{(1)} + \frac{8\,(d-p)\,(p-1)}{\alpha\,(d-1) }\,\phi^\ms{(1)} \\ \nonumber
 & \dub + \frac{4\,(d-p)\,(p-1)}{\alpha^{\,2}\,(d-1) }\,Y^\ms{(1)} \\ \label{TraceAsympEinstein2}
h^{ab}_\ms{(0)} \, \RR_{ab}^\ms{(2)} = & \dub - \left(\frac{(d-1)\,(d-2)}{2 }+ \frac{4\,(p-1)^2}{\alpha^{\,2}\,(d-1) } \right)\,h^\ms{(2)} + \frac{4\,(p-1)^{2}}{\alpha^{\,2}\,(d-1) }\,h_{s}^\ms{(2)} + \frac{8\,(d-p)\,(p-1)}{\alpha\,(d-1) }\,\phi^\ms{(2)} \\ \nonumber
 & \dub + \frac{4\,(d-p)\,(p-1)}{\alpha^{\,2}\,(d-1) }\,Y^\ms{(2)}
\end{align}

Next we consider two components of the Maxwell equation \eqref{MaxwellEOM}. Taking all of the indices to lie in the $p$-dimensional component of $\Sigma$ gives
\begin{align}\label{AsympSphereMaxwell1}
\DD^{a} W_{\alpha a}^\ms{(1)} + 2\,(p-1) \, Z_{\alpha}^\ms{(1)} = & \dub \DD^{a} h_{\alpha a}^\ms{(1)} - \frac{1}{2}\,\DD_{\alpha} h^\ms{(1)} + \DD_{\alpha} h_{s}^\ms{(1)} - \DD^{\beta} h_{\alpha\beta}^\ms{(1)} - 2\, \alpha \, D_{\alpha} \phi^\ms{(1)} \\ \label{AsympSphereMaxwell2}
\DD^{a} W_{\alpha a}^\ms{(2)} + (2\,p-3) \, Z_{\alpha}^\ms{(2)} = & \dub \DD^{a} h_{\alpha a}^\ms{(2)} - \frac{1}{2}\,\DD_{\alpha} h^\ms{(2)} + \DD_{\alpha} h_{s}^\ms{(2)} - \DD^{\beta} h_{\alpha\beta}^\ms{(2)} - 2\, \alpha \, D_{\alpha} \phi^\ms{(2)} ~.
\end{align}
Projecting a single index along the normal vector $n^{a}$ shows that the modes $t_{a_1 \ldots a_{p-1}}^\ms{(n)}$
have vanishing divergence
\begin{align}\label{AsympLongMaxwell1}
\DD^{b} t_{b \, a_{2} \ldots a_{p-1}}^\ms{(1)} = & \dub 0 \\ \label{AsympLongMaxwell2}
\DD^{b} t_{b \, a_{2} \ldots a_{p-1}}^\ms{(2)} = & \dub 0 ~.
\end{align}
The dilaton equation of motion \eqref{DilatonEOM} is straightforward and leads to the following equations
\begin{align}\label{AsympDilaton1}
\DD^{2} \phi^\ms{(1)} - (d-2)\,\phi^\ms{(1)} = & \dub \frac{(d-1)\,(p-1)}{\alpha } \, \left(Y^\ms{(1)} - h_{s}^\ms{(1)} + 2\,\alpha\,\phi^\ms{(1)} + \frac{(d-2)}{2\,(d-1)}\,h^\ms{(1)} \right) \\ \label{AsympDilaton2}
\DD^{2} \phi^\ms{(2)}  = & \dub \frac{(d-1)\,(p-1)}{\alpha } \, \left(Y^\ms{(2)} - h_{s}^\ms{(2)} + 2\,\alpha\,\phi^\ms{(2)} + \frac{1}{2}\,h^\ms{(2)} \right) ~.
\end{align}

In addition to the equations of motion, a few useful results follow from the asymptotic expansion of the Bianchi identities for the field strength and curvature. The twice-contracted Bianchi identity for the curvature is
\begin{equation}
\nabla^{b} R_{ab} - \frac{1}{2}\,\nabla_{a} R = 0 ~.
\end{equation}
The asymptotic expansion of this identity, in conjunction with the equations of motion, implies
\begin{align}
\DD^{\alpha} W_{\alpha a}^\ms{(1)} - \frac{1}{2}\,\DD_{a} Y^\ms{(1)} - (2p-3)\,\delta_{a}{}^{\alpha}\, Z_{\alpha}^\ms{(1)}  = & \dub 0 \\
\DD^{\alpha} W_{\alpha a}^\ms{(2)} - \frac{1}{2}\,\DD_{a} Y^\ms{(2)} - (2p-3)\,\delta_{a}{}^{\alpha}\, Z_{\alpha}^\ms{(2)}  = & \dub 0 ~.
\end{align}
The Bianchi identity for the field strength is
\begin{equation}
\nabla^{a_1} \left((* F)_{a_1 \ldots a_{d+1-p}} \right) = 0
\end{equation}
where `$*$' is the Hodge star operator for the full $d+1$-dimensional metric. When the $d-p$ free indices all lie in the $(d-p)$-dimensional component of $\Sigma$, the asymptotic expansion of this identity gives
\begin{align}
\DD^{\alpha} Z_{\alpha}^\ms{(1)} + \frac{(d-2)}{2}\,Y^\ms{(1)} = & \dub 0 \\
\DD^{\alpha} Z_{\alpha}^\ms{(2)} + \frac{(d-1)}{2}\,Y^\ms{(2)} = & \dub 0 ~.
\end{align}
These equations may also be obtained from the $d$ dimensional Bianchi identities for $\RR_{ab}$ and $\FF_{a_1 \ldots a_{p}}$ on $\Sigma$.

\section{Electric Part of the Weyl Tensor}
\label{sec:Weyl}

The Weyl tensor on $(\MM,g)$ is defined by removing all traces from the Riemann tensor. It is useful to first introduce the Schouten tensor in $d+1$ dimensions
\begin{gather}\label{Schouten}
   S_{ab} = \frac{1}{d-1} \, \left(R_{ab} - \frac{1}{2\,d}\,g_{ab} R \right)~.
\end{gather}
The Weyl tensor can now be written as
\begin{equation}\label{Weyl}
   C^{c}{}_{a d b} = R^{c}{}_{a d b} +
   g^{c}{}_{a} \, S_{b d} - g^{c}{}_{d} \, S_{ab}
   + g_{a d} \, S^{c}{}_{b} -  g_{ab} \, S^{c}{}_{d} \raisebox{14pt}{}~.
\end{equation}
Contracting two indices with the normal vector $n^{a}$ gives the `electric' part of the Weyl tensor
\begin{equation}
 E_{ab} = C_{a c b d} \, n^{c} \, n^{d} ~.
\end{equation}
Notice that $E_{ab}$ is both traceless and orthogonal to $n^{a}$ due to the anti-symmetry properties of the
Weyl tensor.

The electric part of the Weyl tensor is central to our derivation of a defining equation for $\hK_{ab}$. The starting point is a purely geometrical expression for $E_{ab}$ at the constant $\rho$ surface $\Sigma \subset \MM$, given in terms of intrinsic and extrinsic curvatures and their normal derivatives. Contracting two copies of $n^{a}$ into the Weyl tensor \eqref{Weyl} gives
\begin{equation}
  E_{ab} = R_{a c b d} \, n^{c} \, n^{d} + n_{b} \, n^{c} \, S_{a c}
    + n_{a} \, n^{c} \, S_{c b} - S_{ab} - g_{ab} \,n^{c} \, n^{d}
    \, S_{c d} ~.
\end{equation}
Projecting this equation onto $\Sigma$ as in \eqref{projection}, and using the fact that $\perp
E_{ab} = E_{ab}$, we have
\begin{equation}\label{ElectricWeyl1}
  E_{ab} = \perp (R_{a c b d} \,n^{c} \,n^{d}) - \perp S_{ab} -
    h_{ab} \, n^{c} \, n^{d} S_{c d} ~.
\end{equation}
The projections of bulk curvature tensors from appendix \ref{sec:Hyper} can now be used to express the right-hand side of this equation in terms of intrinsic and extrinsic curvatures and their derivatives. The result is
\begin{align}\label{EWeylGeometric}
 E_{ab} = & \left(\frac{d-2}{d-1}\right) \, \left(\frac{1}{d} \, h_{ab}\,\LL_{n} K -
  \LL_{n} K_{ab} \right) + \left(\frac{d-3}{d-1}\right) \, K_{a}{}^{c}\,K_{b c} + \frac{1}{d}\,h_{ab} \, K^{c d} \,K_{c d} \\ \nonumber
 & + \frac{1}{d-1} \, K \, \left(K_{ab} - \frac{1}{d} \, h_{ab} \,K \right) - \frac{1}{d-1} \,
   \left(\RR_{ab} - \frac{1}{d} \, h_{ab} \, \RR \right) \\ \nonumber
 & + \left(\frac{d-2}{d-1}\right) \,  \left(D_{a} a_{b} - \frac{1}{d} \, h_{ab} \,
  D_{c} a^{c}- a_{a}\,a_{b} + \frac{1}{d} \, h_{ab} \, a_{c} a^{c}
  \right) ~.
\end{align}
All of the terms in this expression are manifestly parallel to $\Sigma$, so that $E_{ab}$ is orthogonal to $n^{a}$. It is straightforward to verify that this expression is traceless.

Applying the results of the previous appendix to \eqref{EWeylGeometric}, the asymptotic expansion of the electric part of the Weyl tensor takes the form
\begin{gather}\label{EWeylExpansion}
  E_{ab} = E_{ab}^\ms{(0)} + \rho^{\,2-d}\,E_{ab}^\ms{(1)} + \rho^{\,1-d}\,E_{ab}^\ms{(2)} + \ldots ~.
\end{gather}
The terms in the expansion are given by
\begin{align}\label{E0}
E_{ab}^\ms{(0)} = & \,\,- \frac{1}{d-1}\,\left(\RR_{ab}^\ms{(0)} - \frac{1}{d}\,h_{ab}^\ms{(0)}\,\RR^\ms{(0)} \right) \\
\raisebox{20pt}{}E_{ab}^\ms{(1)} = & \,\,\frac{(d-2)^2}{2\,(d-1)}\,\left( \frac{1}{d}\,h_{ab}^\ms{(0)}\,h^\ms{(1)} - h_{ab}^\ms{(1)} \right) - \frac{1}{d-1}\,\RR_{ab}^\ms{(1)} + \frac{1}{d (d-1)}\,h_{ab}^\ms{(1)}\,\RR^\ms{(0)} \\ \nonumber
	& \, \, +\frac{1}{d (d-1)}\,h_{ab}^\ms{(0)}\,\left( h^{c d}_\ms{(0)}\,\RR_{c d}^\ms{(1)}
	 - h^{c d}_\ms{(1)}\,\RR_{c d}^\ms{(0)} \right)  \\
\raisebox{20pt}{}E_{ab}^\ms{(2)} = & \,\,\frac{(d-1)(d-2)}{2}\,\left( \frac{1}{d}\,h_{ab}^\ms{(0)}\,h^\ms{(2)} -
	h_{ab}^\ms{(2)} \right) - \frac{1}{d-1}\,\RR_{ab}^\ms{(2)} + \frac{1}{d (d-1)}\,h_{ab}^\ms{(2)}
	\,\RR^\ms{(0)} \\ \nonumber
	& \, \, +\frac{1}{d (d-1)}\,h_{ab}^\ms{(0)}\,\left( h^{c d}_\ms{(0)}\,\RR_{c d}^\ms{(2)}
	 - h^{c d}_\ms{(2)}\,\RR_{c d}^\ms{(0)} \right)
\end{align}
These expressions depend only on the geometric result \eqref{EWeylGeometric} and the asymptotic form of the metric \eqref{metricBC}.

In AF gravity the leading term in the expansion \eqref{EWeylExpansion} vanishes, but this is not the case for a theory with ALD boundary conditions.
\begin{align}
   E_{ij}^\ms{(0)} = & \,\, - \frac{4\,p\,(p-1)}{d\,\alpha^{2} }\,h_{ij}^\ms{(0)} \\
   E_{\alpha\beta}^\ms{(0)} = & \,\, \frac{4\,(d-p)\,(p-1)}{d\,\alpha^{2} }\,h_{\alpha\beta}^\ms{(0)} ~.
\end{align}
This is due to the presence of a p-form flux at spatial infinity, which, as is clear from the constraints \eqref{Ricci1} and \eqref{Ricci2}, is only possible if $\RR_{ab}^\ms{(0)}$ is not traceless.
One consequence of this result is that the defining equation \eqref{DefiningEquation} receives contributions that were not present in AF gravity \cite{Mann:2005yr, Mann:2006bd, Mann:2008ay}.


\end{document}